\documentclass[11pt, a4paper]{article}
\pdfoutput=1

\usepackage{amssymb,amsmath,amsfonts, mathtools, mathrsfs, graphicx, caption, subcaption, doi, cite, hyperref, url, dsfont, enumerate}

\usepackage[utf8]{inputenc} 
\usepackage[dvipsnames]{xcolor}
\usepackage{amsmath}

\usepackage{verbatim}
\usepackage{amsfonts,euscript,amssymb,stmaryrd,braket}
\usepackage{tikz}
\usetikzlibrary{arrows,decorations.markings,patterns}
\usepackage{slashed}

\def\clock{{\count0=\time
		\divide\count0 60
		\ifnum\count0<10 0\fi\the\count0
		\multiply\count0 -60 \advance\count0 \time
		:\ifnum\count0<10 0\fi \the\count0
}}
\newcommand{\timestamp}{{\small\vbox{\hbox{\tt\jobname.tex}
			\hbox{\the\day/\the\month/\the\year, \clock}}}}

\makeatletter
\let\old@startsection=\@startsection
\let\oldl@section=\l@section
\renewcommand{\@startsection}[6]{\old@startsection{#1}{#2}{#3}{#4}{#5}{#6\mathversion{bold}}}
\renewcommand{\l@section}[2]{\oldl@section{\mathversion{bold}#1}{#2}}
\makeatother

\numberwithin{equation}{section}

\graphicspath{{figures/}}

\begin{document}
	\renewcommand{\thefootnote}{\arabic{footnote}}
		
	\overfullrule=0pt
	\parskip=2pt
	\parindent=12pt
	\headheight=0in \headsep=0in \topmargin=0in \oddsidemargin=0in
		
	\vspace{ -3cm} \thispagestyle{empty} \vspace{-1cm}
	\begin{flushright} 
		USTC-ICTS/PCFT-26-40
	\end{flushright}

\begin{center}
	\vspace{1.2cm}
	{\Large\bf \mathversion{bold}
	Effective dynamics and quantum information in\\ de Sitter wedge holography }
	\\
	\vspace{.2cm}
	\noindent
	{\Large\bf \mathversion{bold}}
	
	\vspace{0.4cm} {
		Sabyasachi Maulik$^{\,a, b}$\footnote[1]{mauliks@ustc.edu.cn} and Soumen Pari$^{\,c, b}$\footnote[2]{soumen.pari@saha.ac.in}},
	\vskip  0.4cm
	
	\small
	{\em
		$^{a}\,$Interdisciplinary Center for Theoretical Study, University of Science and Technology of China, Hefei, Anhui 230026, China.
		\vskip 0.1cm
        $^{b}\,$Peng Huanwu Center for Fundamental Theory, Hefei, Anhui 230026, China.
        \vskip 0.1 cm
		$^{c}\,$Theory division, Saha Institute of Nuclear Physics, 1/AF, Bidhannagar, West Bengal 700064, India.
		\vskip 0.1cm
		$^{d}\,$ Homi Bhabha National Institute, Training School Complex, Anushaktinagar,\\ Mumbai 400094, India.
	}
	\normalsize
	
\end{center}

\vspace{0.1cm}

\begin{abstract}
In this paper, we study codimension-two holography in a de Sitter (dS) wedge setup, based on the idea of wedge holography. We consider a $d+1$-dimensional Anti-de Sitter (AdS) bulk spacetime bounded by two end-of-the-world branes with $d$-dimensional de Sitter geometry. We propose that this configuration is holographically dual to a conformal field theory (CFT) living on a $d-1$-dimensional sphere. Our computations of the partition function and holographic entanglement entropy support this duality and indicate that the dual CFT is non-unitary. We also analyze the mass spectrum in dS wedge holography. We verify the first law of entanglement entropy within this framework. Finally, we make use of the island prescription to study the Page curve in a simplified model within our dS wedge holography framework.
\end{abstract}

\tableofcontents
\section{Introduction} \label{intro}

The AdS/CFT correspondence \cite{Maldacena:1997re, Gubser:1998bc, Witten:1998qj} provides a powerful framework for studying quantum gravity in terms of an ordinary quantum field theory. Apart from giving a non-perturbative description of gravity in anti-de Sitter spacetime, it has also revealed a close relation between spacetime geometry and quantum information. A central example is the Ryu-Takayanagi prescription and its covariant extension \cite{Ryu:2006bv, Ryu:2006ef, Hubeny:2007xt}, which express the entanglement entropy of a boundary subregion in terms of the area of a codimension-two bulk extremal hypersurface. These ideas have led to a better understanding of bulk reconstruction and the emergence of gravitational dynamics from entanglement \cite{Hamilton:2006az, Lashkari:2013koa, Faulkner:2013ica, Swingle:2014uza, Dong:2016eik, Faulkner:2017tkh}.

Wedge holography is a recent extension of the usual holographic
correspondence. In its simplest form, one considers a region of
AdS$_{d+1}$ spacetime bounded by two end-of-the-world (ETW) branes. Gravity in this wedge is proposed to be dual to quantum gravity on the two $d$-dimensional branes, and also to a CFT$_{d-1}$ living on their codimension-two intersection
\cite{Akal:2020wfl, Miao:2020oey, Miao:2021ual, Geng:2022tfc, Geng:2022slq, Ogawa:2022fhy, Yadav:2023qfg}. Thus, wedge holography gives three related descriptions of the same system
\begin{align*}
    \text{Classical gravity in the wedge} &\cong
    \text{Quantum gravity on the branes} \\
    &\cong \text{CFT on the codimension-two defect}
\end{align*}
Several checks of this proposal have been performed. These include computations of the free energy, Weyl anomaly, entanglement entropy, and correlation functions. The effective gravitational action and the spectrum of fluctuations on the branes have also been studied \cite{Akal:2020wfl, Miao:2020oey, Ogawa:2022fhy}. In particular, Neumann boundary conditions on both branes allow a massless graviton mode \cite{Miao:2020oey}, and the first law of entanglement entropy has been verified in this setting \cite{Hu:2022lxl}. Further, wedge holography has been extended to flat spacetime, where it may have a relation to celestial holography \cite{Ogawa:2022fhy}. It has also provided a useful setting for studying entanglement islands and Page curves in theories with dynamical gravity \cite{Miao:2023unv, Aguilar-Gutierrez:2023tic}.

It is natural to ask whether a similar construction can be used when the induced geometry on the branes is de Sitter (dS) spacetime. This question is interesting because holography in de Sitter spacetime \cite{Strominger:2001pn, Maldacena:2002vr} is less well understood than holography in AdS. A de Sitter wedge can be constructed by foliating AdS$_{d+1}$ with dS$_{d}$ slices and restricting the radial coordinate between two ETW branes. An earlier study of this geometry described the two branes as entangled de Sitter universes. It also studied the Page curve, holographic complexity, and the appearance of de Sitter Jackiw--Teitelboim (JT) gravity from brane fluctuations \cite{Aguilar-Gutierrez:2023tic}. These results suggest that de Sitter wedge holography is a useful laboratory for studying both de Sitter holography and quantum information quantities.

The purpose of the present work is to examine the basic holographic dictionary of the de Sitter wedge in more detail. We consider an $\mathrm{AdS}_{d+1}$ spacetime bounded by two ETW branes. The induced geometry on each brane is a $\mathrm{dS}_{d}$ spacetime. We propose that this system is dual to a $\mathrm{CFT}_{d-1}$ defined on the spherical codimension-two defects situated at the intersection of the branes. We test this proposal using several quantities that probe different parts of
the correspondence.

First, we integrate over the radial direction of the wedge and show that the bulk gravitational action reduces to an Einstein gravity action on the de Sitter branes, with an effective Newton constant determined by the brane
positions. We then evaluate the gravitational partition function for different spacetime dimensions. The universal terms give imaginary central charges, as expected for a non-unitary CFT appearing in the dS/CFT correspondence \cite{Strominger:2001pn, Maldacena:2002vr}. We independently obtain the same central charges by computing the holographic entanglement entropy. We further study the first law of entanglement \cite{Blanco:2013joa, Wong:2013gua, Bhattacharya:2012mi, Allahbakhshi:2013rda, Hu:2022lxl} for graviton fluctuations in the $\left(d+1 \right)$-dimensional AdS background.

The rest of the paper is organized as follows: In Section \ref{sec2}, we introduce the de Sitter wedge geometry, derive the effective action on the branes, evaluate the partition function, and study the graviton mass spectrum. In Section \ref{sec:HEE}, we compute the holographic entanglement entropy, verify its first law for massive and massless fluctuations, and study the timelike entanglement entropy between the two defects. In Section \ref{sec:Page_curve}, we study the appearance of islands and obtain the Page curve for entanglement entropy. We summarize our results and conclude in Section \ref{sec:conclusion}. Some technical details are
collected in the appendices.

\section{de Sitter wedge holography } \label{sec2}

In this section, we begin our exploration of dS wedge holography. We first describe the geometric setting where a pair of end-of-the-world (ETW) dS$_{d}$ branes are embedded in an AdS$_{d+1}$ bulk spacetime. We observe that the on-shell gravitational action in $\left(d + 1 \right)$-dimensional bulk AdS spacetime is equal to the induced gravity action on the dS$_{d}$ branes -- leading to a realisation of braneworld dS$_{d}$/CFT$_{d-1}$ correspondence. We compute the central charge of the dual CFT from the Euclidean gravitational partition function. Next, we study the mass spectrum of gravitons on the branes with both Dirichlet and Neumann boundary conditions.

\begin{figure}[t]
    \centering 
    \includegraphics[width = 0.5\linewidth]{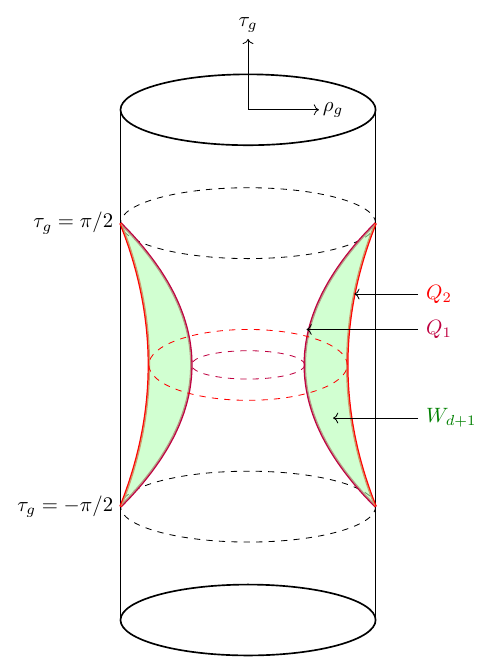} 
   \caption{The geometry for dS wedge holography. $Q_1$ (purple) and $Q_2$ (red) are two ETW branes embedded in $AdS_{d+1}$ space with global time and radial coordinates $\tau_g$ and $\rho_g$ respectively. The branes extend between $\tau_g = \pm\frac{\pi}{2}$, ending on a pair of co-dimension-2 Euclidean defects. The shaded region between the branes is denoted as the wedge $W_{d+1}$.}
    \label{fig:dS_wedge_geometry}
\end{figure}

The setup is based on the geometric construction of dS wedge holography, borrowed from \cite{Miao:2020oey, Ogawa:2022fhy, Aguilar-Gutierrez:2023tic} and illustrated in Figure \ref{fig:dS_wedge_geometry}. Let $W_{d+1}$ be the bulk
wedge space within an asymptotically anti-de Sitter geometry, bounded by two ETW branes $Q=Q_{1}\cup Q_{2}$. The induced geometry on each brane is $d$-dimensional de Sitter spacetime. The formulation of dS wedge holography on the branes is described by the action
\begin{equation}\label{action}
	I=\int_{W} d^{d+1}x \sqrt{-g} \left(\hat{R}_{W}-2\Lambda_{d+1}\right) + \sum_{a = 1}^{2} \int_{Q_{a}} d^{d}x \sqrt{-h^{(a)}}\left(\mathcal{K}^{(a)} - T^{(a)}\right),
\end{equation}
where $\hat{R}_{W}$ denotes the Ricci scalar of the $\left(d + 1 \right)$-dimensional bulk spacetime, and $\mathcal{K}^{(a)}$ represents the extrinsic curvature of each of the $d$-dimensional ETW brane. The quantity $h^{(a)}_{ij}$ correspond to the induced metric on each brane, and the parameter $T^{(a)}$ refers to the brane-tension for $a=1,2$. In this article, we consider $h^{(1,\,2)}_{ij}$ to be the metric of $d$-dimensional de Sitter spacetime. For simplicity, we set the Newton's constant $16 \pi G_{N}$ and the AdS radius $L$ to unity.

To formalize the construction, we consider a foliation of the bulk manifold with dS$_{d}$ slices. The AdS$_{d+1}$ metric with de Sitter slicing can be written as \cite{Miao:2020oey} 
\begin{equation} \label{dS slicing metric}
	ds^{2}_{d+1} = d\rho^{2} + \sinh^{2} \rho\, h_{ij}(x)dx^{i}dx^{j} \, ,
\end{equation}
where $h_{ij}(x)$ denotes the $d$-dimensional de Sitter line element with unit curvature radius in any convenient coordinate system. E.g. if we use global coordinates for the de Sitter slices, the complete metric is expressed as
\begin{equation} \label{Ads_global_ds_slice_metric}
    ds^{2}_{d+1} = d\rho^2 + \sinh^2\rho \left(-d\tau^2 + \cosh^2\tau\,  d\Omega_{d-1}^2 \right),
\end{equation}
where $d\Omega_{d-1}^{2}$ denotes the metric on $\left(d-1 \right)$-dimensional sphere $S^{d-1}$.

The wedge geometry $W_{d+1}$ is defined by restricting the radial coordinate within the range  
\begin{equation}
	\rho_{1} \leq \rho \leq \rho_{2}.
\end{equation}
Therefore, $\rho_{1}$ and $\rho_{2}$ denote the positions of the ETW branes $Q_{1}$ and $Q_{2}$, respectively. In general, the ``UV brane" $Q_{2}$ is located close to the AdS conformal boundary such that the induced gravity on $Q_{1}$ remains weak and approximately local. This placement requires the brane tension to be positive, $T_{2} > 0$. On the other hand, the ``IR brane" $Q_{1}$ is placed in the interior, where the induced gravity behaves in a non-local manner, similar to the original setup of \cite{Randall:1999ee,Randall:1999vf} . To sustain a positive curvature on $\rho_{1}$, one typically needs a negative brane tension, $T_{1} < 0$ \cite{Aguilar-Gutierrez:2023tic}.  

The bulk regions $\rho < \rho_{1}$ and $ \rho > \rho_{2}$ are integrated out through a Wilson-type procedure \cite{Heemskerk:2010hk, Faulkner:2010jy, Guijosa:2022jdo}, leading to the effective boundary of the wedge geometry $\partial W_{d+1} = \rho_{1} \cup \rho_{2}$. The coarse-graining generates induced gravity theories on the branes, which take the form of Einstein gravity supplemented by an infinite series of higher-derivative corrections. These corrections are suppressed as the branes approach the AdS boundary \cite{deHaro:2000vlm,Balasubramanian:1999re,Emparan:1999pm}.

For the consistency of the construction, the branes must satisfy the Israel junction conditions \cite{Israel:1966rt}, which take the following form\cite{Miao:2023unv}
\begin{equation}\label{NBC}
	\mathcal{K}^{(a)}_{ij}-\left(\mathcal{K}^{(a)} -T^{(a)}\right) h^{(a)}_{ij}=0.
\end{equation}

\subsection{Effective action}

In the wedge holography framework, the central statement is that, in the large $N$ limit, the partition function of the $(d-1)$-dimensional conformal field theory is determined by the classical Euclidean gravitational action evaluated on the $(d+1)$-dimensional asymptotically AdS spacetime containing a wedge \cite{Akal:2020wfl, Miao:2020oey, Miao:2023unv}. Explicitly, this relation is expressed as
\begin{equation}
    Z_{\mathrm{CFT}_{d-1}} = e^{-I_{\mathrm{AdS}_{d+1}}}.
\end{equation}
We now demonstrate that wedge holography, described by AdSW$_{d+1}$/CFT$_{d-1}$ with the solution given in \eqref{dS slicing metric}, is equivalent to dS$_{d}$/CFT$_{d-1}$ correspondence with Einstein gravity on the brane. In the dS$_{d}$/CFT$_{d-1}$ correspondence, we assume the relation \cite{Strominger:2001pn, Maldacena:2002vr}
\begin{equation}
    Z_{\mathrm{CFT}_{d-1}} = e^{-I_{\mathrm{dS}_{d}}},
\end{equation}
where $I_{\mathrm{dS}_{d}}$ refers to the Euclidean gravitational action on the ETW branes. Therefore, the equivalence between wedge holography and dS/CFT correspondence can be established by showing that the two gravitational actions coincide
\begin{equation}
    I_{\mathrm{AdS}_{d+1}} = I_{\mathrm{dS}_{d}}.
\end{equation}
We substitute the metric \eqref{dS slicing metric} into the action \eqref{action}, perform the analytic continuation $t \to -i \tau$, and impose the Neumann boundary condition \eqref{NBC}  together with the expression for the Ricci scalar in \eqref{Ricci scalar}. The resulting on-shell Euclidean action is given by
\begin{equation}
    \begin{aligned}
        I_{\mathrm{AdS}_{d+1}} &= \frac{1}{16\pi G_N}
        \int_{\rho_1}^{\rho_2} d\rho\, \sinh^{d}\rho \int_{Q_1 \cup Q_2} d^{d}x\, \sqrt{h} \left(R_h \mathrm{csch}^{2} \rho - d \left(2 + \left(d - 1 \right) \coth^{2} \rho \right) + d \left(d - 1 \right) \right) \\ & \hspace{4 em} + \frac{1}{8\pi G_N} \left(\int_{Q_1} d^{d}x\, \sqrt{h}\, \sinh^{d}\rho_1 \coth\rho_1 - \int_{Q_2} d^{d}x\, \sqrt{h}\, \sinh^{d}\rho_2 \coth\rho_2 \right),
    \end{aligned}
\end{equation}
where $R_{h}$ is the Ricci scalar on each dS$_{d}$ brane. Integrating over the AdS$_{d+1}$ radial direction $\rho$, we obtain
\begin{equation}
    I_{\mathrm{AdS}_{d+1}} = \frac{1}{16\pi G_N^{(d)}} \int_{\mathcal Q_1 \cup \mathcal Q_2} \sqrt{h}\, \left(R_h - (d-1)(d-2) \right),
\end{equation}
which is equal to the on-shell Euclidean gravitational action $I_{dS_{d}}$ on the brane, with the $d$-dimensional Newton's constant given by
\begin{equation}
    \frac{1}{G^{(d)}_{N}}=\frac{1}{G_{N}}\int_{\rho_{1}}^{\rho_{2}}\sinh^{d-2}\rho \ d\rho
\end{equation}
 It is important to emphasize that, in the above derivation, the induced metric $h$ is treated off-shell. In other words, we do not impose Einstein's equations on the brane $Q_a$ $\left(a = 1, 2 \right)$. Further, we made use of the relations $-2\Lambda = d(d-1)$, and $\mathcal{K} - T = \coth \rho$, together with the following integral identity
\begin{equation}
    \int_{\rho_1}^{\rho_2} \sinh^{d-2}\rho \, \bigl(d-1 + d\sinh^2\rho\bigr)\, d\rho
    = \sinh^d \rho_2 \coth \rho_2 - \sinh^d \rho_1 \coth \rho_1.
\end{equation}
\subsection{Partition function  }
Having established the effective gravitational description, we now evaluate the Euclidean on-shell action and extract the corresponding partition function of the dual conformal field theory.The gravity action on the wedge region is given by
\begin{align}
I_{G}
=&\;
\frac{1}{16\pi G_{N}}
\int_{W} \sqrt{-g}\,
\left( \hat{R}_W - 2\Lambda_{d+1} \right)
\nonumber\\[6pt]
&\;
-\frac{1}{8\pi G_{N}}
\Bigg[
\int_{Q^{(1)}} \sqrt{-h}\,
\left( \mathcal{K}^{(1)} - T^{(1)} \right)
-
\int_{Q^{(2)}} \sqrt{-h}\,
\left(\mathcal{K} ^{(2)} - T^{(2)} \right)
\Bigg].
\end{align}

To regulate the divergent spacetime volume, we restrict the spacetime to the region $0\leq \tau \leq \tau_{\infty}$. Using
\begin{equation}
    \int_{Q^{(i)}}\sqrt{-h}
    =
    \sinh^{d}\rho_i\,\omega_{d-1}
    \int_{0}^{\tau_{\infty}}
    d\tau\,\cosh^{d-1}\tau,
\end{equation}
where $\omega_{d-1}$ denotes the volume of the unit $(d-1)$-sphere, and defining
\begin{equation}
    J_d
    =
    \int_{0}^{\tau_{\infty}}
    d\tau\,\cosh^d\tau,
\end{equation}
the on-shell action can be written as
\begin{equation}
    I_G
    =
    \frac{(d-1)\omega_{d-1}J_{d-1}}
    {8\pi G_N}
    \int_{\rho_1}^{\rho_2}
    \sinh^{d-2}\rho\,d\rho.
\end{equation}
The integral $J_d$ satisfies a simple recursion relation. Integrating by parts yields
\begin{equation}
J_d
=
\frac{1}{d}
\cosh^{d-1}\tau_{\infty}
\sinh\tau_{\infty}
+
\frac{d-1}{d}
J_{d-2}.
\end{equation}
In the following, we evaluate the on-shell action explicitly for $d=3$, $4$, and $5$, and compare the resulting partition functions with the expected ultraviolet structure of the corresponding dual conformal field theories.

\subsubsection*{$d=3$ Case}

For $d=3,$ the on-shell action becomes
\begin{equation}
    I_{G}=\frac{\left(\cosh\rho_{2}-\cosh{\rho_{1}}\right)}{16 G_{N}}\left(e^{2\tau_{\infty}}+4 \tau_{\infty}\right).
\end{equation}
Identifying the de Sitter cutoff $\tau_{\infty}$ with the ultraviolet cutoff $\epsilon$ of the dual two-dimensional CFT on $S^2$ through
\begin{equation}\label{geometric cutoff}
    \epsilon=e^{-\tau_{\infty}},
\end{equation}
the on-shell action becomes
\begin{equation}
    I_{G}
    =
    \frac{(\cosh\rho_{2}-\cosh\rho_{1})}
    {8G_{N}\epsilon^{2}}
    -
    \frac{(\cosh\rho_{2}-\cosh\rho_{1})}
    {2G_{N}}
    \log\epsilon.
\end{equation}
The above result can be compared with the sphere partition function of a two-dimensional CFT with central charge $c$, which takes the form \cite{Duff:1993wm,Henningson:1998gx}
\begin{equation}
    Z_{\mathrm{CFT}}
    \sim
    \exp\left(
    \frac{A}{\epsilon^{2}}
    -
    \frac{c}{3}\log\epsilon
    \right),
\end{equation}
where $A$ is a non-universal constant, whereas the logarithmic term is universal and is determined by the conformal anomaly. Identifying the  partition function as $Z_{\mathrm{CFT}}=e^{iI_G}$, we obtain the central charge
\begin{equation}\label{central cahrge}
    c
    =
    i\,\frac{3}{2G_N}
    \left(
    \cosh\rho_2-\cosh\rho_1
    \right).
\end{equation}

\subsubsection*{$d=4$ Case}

For $d=4$, the on-shell action is found to be 
\begin{equation}
    I_{G}=\frac{3\pi}{8 G_{N}}\left((\sinh2\rho_{2}-\sinh 2 \rho_{1})-2(\rho_{2}-\rho_{1})\right)\left(\frac{1}{24}e^{3\tau_{\infty}}+\frac{1}{3}e^{\tau_{\infty}}\right)
\end{equation}
Using the cutoff relation \eqref{geometric cutoff},this expression becomes
\begin{equation}
    I_{G}=\frac{\pi\left((\sinh2\rho_{2}-\sinh 2 \rho_{1})-2(\rho_{2}-\rho_{1})\right)}{64 G_{N}\epsilon^3}+\frac{\pi\left((\sinh2\rho_{2}-\sinh 2 \rho_{1})-2(\rho_{2}-\rho_{1})\right)}{8 G_{N}\epsilon}.
\end{equation}
As expected for an odd-dimensional boundary conformal field theory, the logarithmic divergence is absent, reflecting the absence of a conformal anomaly.

\subsubsection*{$d=5$ Case}

For $d=5$,we can evaluate the on-shell action as below
\begin{equation}
    I_{G}=\frac{\pi}{48G_{N}}\left(\frac{1}{12}e^{4\tau_{\infty}}+\frac{2}{3}e^{\tau_{\infty}}+2\tau_{\infty}\right)\left((\cosh 3\rho_{2}-\cosh 3\rho_{1})-9(\cosh\rho2-\cosh\rho_{1})\right)
\end{equation}
In terms of $CFT$ cut off \eqref{geometric cutoff}, we find 
\begin{equation}
     I_{G}=\frac{\pi}{6G_{N}}\left(\frac{1}{96\epsilon^4}+\frac{1}{12\epsilon^2}-\frac{1}{4}\log\epsilon\right)\left((\cosh 3\rho_{2}-\cosh 3\rho_{1})-9(\cosh\rho2-\cosh\rho_{1})\right).
\end{equation}
The logarithmic divergence of the on-shell action can be compared with the universal contribution to the partition function of a four-dimensional conformal field theory on $S^4$ with central charge a, c. The latter satisfies \cite{Duff:1993wm,Henningson:1998gx}
\begin{equation}\label{4 sphere partition fn}
\begin{aligned}
\epsilon \frac{d}{d\epsilon} \log Z_{\text{CFT}}
&= -\frac{1}{2\pi}
\left\langle
\int d^4 x \sqrt{g}\, T^{\mu}{}_{\mu}
\right\rangle_{S^4}
\\[4pt]
&= -\frac{1}{2\pi}
\int d^4 x \sqrt{g}
\left(
\frac{a}{8\pi}\,
\widetilde{R}_{\mu\nu\rho\sigma}
\widetilde{R}^{\mu\nu\rho\sigma}
-
\frac{c}{8\pi}\,
W_{\mu\nu\rho\sigma}
W^{\mu\nu\rho\sigma}
\right)
\\[4pt]
&= -2a\,\chi(S^4)
\\[4pt]
&= -4a .
\end{aligned}
\end{equation}
In deriving the third line, we have employed the trace anomaly
\begin{equation}
    \left\langle T^{\mu}{}_{\mu}\right\rangle
    =
    \frac{a}{8\pi}
    \widetilde{R}_{\mu\nu\rho\sigma}
    R^{\mu\nu\rho\sigma}
    -
    \frac{c}{8\pi}
    W_{\mu\nu\rho\sigma}
    W^{\mu\nu\rho\sigma},
\end{equation}
while the last equality follows from the Gauss--Bonnet theorem,
\begin{equation}
    \frac{1}{32\pi^2}
    \int d^4x\,\sqrt{g}\,
    \widetilde{R}_{\mu\nu\rho\sigma}
    R^{\mu\nu\rho\sigma}
    =
    \chi(M),
\end{equation}
together with the fact that the Weyl tensor vanishes on $S^4$. Integrating Equation \eqref{4 sphere partition fn}, the universal logarithmic contribution to the partition function is
\begin{equation}
    \log Z_{\mathrm{CFT}}
    =
    4a\log\epsilon
    +\text{(non-universal terms)}.
\end{equation}
Finally, identifying the partition function as $Z_{\mathrm{CFT}}=e^{iI_G}$, we obtain
\begin{equation}\label{4d CFT central cgarhe}
    a
    =
    -i\,
    \frac{\pi}{96G_N}
    \left[
    (\cosh3\rho_2-\cosh3\rho_1)
    -
    9(\cosh\rho_2-\cosh\rho_1)
    \right].
\end{equation}

\subsection{Mass spectrum of gravitons}

In wedge holography, gravity on the brane is induced from the higher-dimensional bulk theory. Therefore, an important question is whether the resulting lower-dimensional theory describes localized Einstein gravity. A natural diagnostic of this phenomenon is provided by the spectrum of graviton fluctuations.

In this subsection, we study the graviton mass spectrum on the brane. The presence of a normalizable massless spin-2 mode indicates that gravity is localized on the brane, whereas the massive Kaluza–Klein modes encode corrections to the effective low-energy gravitational dynamics. Since graviton fluctuations are directly governed by the background geometry and the boundary conditions, their spectrum provides a robust probe of the induced gravitational theory. Analysing the spectrum under Dirichlet and Neumann boundary conditions therefore allows us to investigate the emergence and consistency of localized gravity in wedge holography.
 
 We focus on dimension $d+1 = 4$. We adopt the following ansatz for the perturbed metric and the embedding profile of the branes $Q^{(a)}$:
  \begin{align}
      &ds^2 = d\rho^2+\sinh^2(\rho)\left(h^{(0)}_{ij}(x)+\epsilon H(\rho)h^{(1)}_{ij}(x)\right)dx^{i}dx^{j}+\mathcal{O}(\epsilon^2), \label{perturbed metric}\\
      &Q_{1} : \rho=\rho_{1} + \mathcal{O}(\epsilon^2) ,
      \qquad Q_{2} : \rho=\rho_{2} + \mathcal{O}(\epsilon^2),
\end{align}
where $h^{(0)}_{ij}$ is the de Sitter metric with unit radius, and $h^{(1)}_{ij}$ is the perturbation. The parameter $\epsilon$ controls the order of the perturbative expansion. In terms of the bulk metric perturbation, we have
\begin{equation}
    \delta g_{\rho\mu}=0,\qquad \delta g_{ij}=\sinh^2(\rho)H(\rho)h^{(1)}_{ij}(x).
\end{equation}
Imposing the transverse-traceless (TT) gauge
\begin{equation}
    \nabla^{\mu}\delta g_{\mu\nu}=0,\quad g^{\mu\nu}\delta g_{\mu\nu}=0,
\end{equation}
we obtain
\begin{equation}\label{TT gauge}
    D^{i}h^{(1)}_{ij}=0,\quad h^{(0)ij}h^{(1)}_{ij}=0,
\end{equation}
where $\nabla_{\mu}$ and $D_i$ denote the covariant derivatives associated with the bulk metric $g_{\mu\nu}$ and the background metric $h^{(0)}_{ij}$, respectively. Substituting \eqref{perturbed metric} together with the TT gauge condition \eqref{TT gauge} into the Einstein equations, and performing a separation of variables, we obtain
\begin{align}
    \left(\Box -2 \right) h^{(1)}_{ij}(x) &= m^{2} h^{(1)}_{ij}(x), \label{brane equation}\\
   \sinh^{2}(\rho) H''(\rho) + 3\sinh\rho\cosh\rho\, H'(\rho) &= - m^{2}H(\rho), \label{radial eqn}
\end{align}
here $m$ represents the mass of the graviton, and $\Box = D_k D^k$ denotes the d'Alembertian operator constructed from the background metric $h^{(0)}_{ij}$. The general solution to equation \eqref{radial eqn} is given by
\begin{align} \label{eq:final mass spectrum equation}
H(\rho) = \text{csch}(\rho) \left(c_{1}P^{\lambda}_{1} \left(\cosh \rho \right) + c_{2} Q^{\lambda}_{1} \left(\cosh \rho \right) \right),
\end{align}
where $P_{1}^{\lambda}$ and $Q_{1}^{\lambda}$ are the Legendre polynomials, and $\lambda$ is given by
\begin{equation}\label{lambda of mass spectrum equation}
    \lambda=\sqrt{1- \  m^2}.
\end{equation}
We will now check the solution with Dirichlet and Neumann boundary conditions.

\subsubsection*{Dirichlet Boundary condition}

\begin{figure}[t]
    \centering
    \begin{subfigure}[t]{0.48\textwidth}
        \centering
        \includegraphics[width=\textwidth]{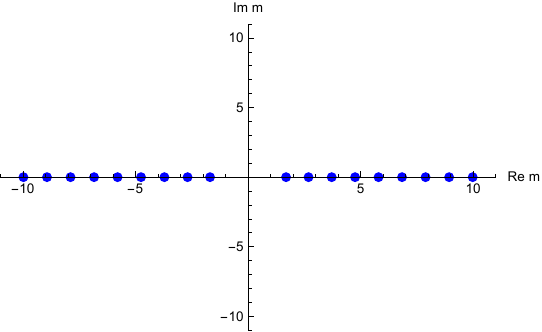}
        \label{fig:left-Dirichlet 3d mass spectrum}
    \end{subfigure}
    \hfill
    \begin{subfigure}[t]{0.48\textwidth}
        \centering
        \includegraphics[width=\textwidth]{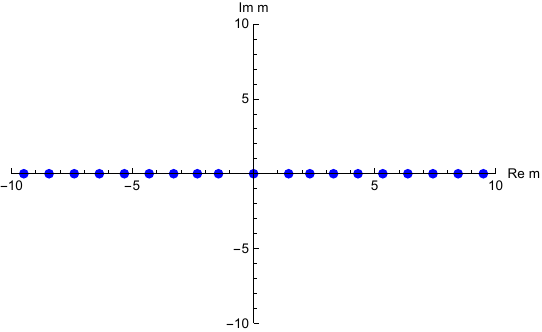}
        \label{fig:right-Dirichlet 3d mass spectrum}
    \end{subfigure}
    \caption{Plots of the roots of the Dirichlet \eqref{eq:DBC equation mass spectrum} (left) and Neumann \eqref{eq:NBC equation mass spectrum} (right) boundary conditions as functions of $m$ for $\rho_{1}=0.1$ and $\rho_{2}=10$.}
    \label{fig:Dirichlet 3d mass spectrum}
\end{figure}

Using equation \eqref{eq:final mass spectrum equation}, the Dirichlet boundary condition for the mass spectrum reads
\begin{equation}\label{eq:BBC con}
    H(\rho_{1})=0,\quad H(\rho_{2})=0.
\end{equation}
Equivalently, we look for values of $m$ that are a solution to
\begin{equation}\label{eq:DBC equation mass spectrum}
    D^{\mathrm{dS}} \left(m, \rho_{1}, \rho_{2} \right) = P_{\lambda}^{1} \left(\cosh\rho_{1} \right) Q_{\lambda}^{1} \left(\cosh\rho_{2} \right) - P_{\lambda}^{1} \left(\cosh\rho_{2} \right) Q_{\lambda}^{1} \left(\cosh\rho_{1} \right) = 0,
\end{equation}
where $\lambda=\sqrt{1-m^2}$ as given in equation \eqref{lambda of mass spectrum equation}. The equation can be analysed using numerical methods. We find that there are an infinite number of solutions for discrete real values of $m$. Our observations are recorded in Figure \eqref{fig:Dirichlet 3d mass spectrum} (Left). Importantly, $m = 0$ does not appear to satisfy the Dirichlet boundary condition; forcing us to conclude that it is impossible to obtain localised gravity on the dS$_{d}$ branes with Dirichlet b.c.

\subsubsection*{Neumann Boundary condition}

Next we consider the situation where we impose Neumann Boundary condition on the two ETW dS$_{d}$ branes
\begin{equation}\label{eq:NBC con}
    \partial_{\rho}H \left(\rho_{1} \right) = 0,\qquad \partial_{\rho}H \left(\rho_{2} \right) = 0,
\end{equation}
 To simplify the equations, we use the recurrence relation of the associated Legendre function
\begin{equation}
    \left(x^2 - 1 \right) \frac{d}{dx} P_{\nu}^{\lambda}(x) = -(\nu+1) x P_{\nu}^{\lambda}(x) + \left(\nu - \lambda+1 \right) P_{\nu+1}^{\lambda}.
\end{equation}
The same relation also holds for $Q_{\nu}^{\lambda}$. For the present problem, $x=\cosh\rho$, $\nu=1$, and $\lambda = \sqrt{1-m^2}$. After some algebra, we find that the Neumann boundary condition is equivalent to
\begin{equation} \label{eq:NBC equation mass spectrum}
    \begin{split}
        \left| N^{\mathrm{dS}} \left(m,\rho_{1}, \rho_{2} \right) \right| &= \left(\left(3\cosh\rho_{1}\, P_{1}^{\lambda} \left(\cosh\rho_{1} \right) + \left( \lambda-2 \right) P_{2}^{\lambda} \left(\cosh\rho_{1} \right) \right) \right. \\
        &\left. \hspace{2 em} \times \left(3\cosh\rho_{2}\,Q_{1}^{\lambda} \left(\cosh\rho_{2} \right) + \left( \lambda -2\right) Q_{2}^{\lambda} \left(\cosh\rho_{2} \right) \right) \right)\\ 
        &\hspace{3 em} - \left( \left(3\cosh{\rho_{2}}\, P_{1}^{\lambda} \left(\cosh\rho_{2} \right) +\left( \lambda -2\right) P_{2}^{\lambda} \left(\cosh{\rho_{2}} \right) \right) \right.\\
        &\left. \hspace{4 em} \times \left(3\cosh{\rho_{1}}\, Q_{1}^{\lambda} \left(\cosh\rho_{1} \right) + \left(\lambda -2\right)\, Q_{2}^{\lambda} \left(\cosh{\rho_{1}} \right) \right) \right) = 0.
    \end{split}
\end{equation}
We can study the roots of the last equation numerically. We find an infinite tower of discrete solutions for for real values of $m$. In addition, unlike the Dirichlet case, the Neumann boundary condition admits a massless mode corresponding to $m=0$. The resulting mass spectrum is shown in the right panel of Figure \ref{fig:Dirichlet 3d mass spectrum}.
\begin{figure}[t]
    \centering
    \begin{subfigure}[t]{0.48\textwidth}
        \centering
        \includegraphics[width=\textwidth]{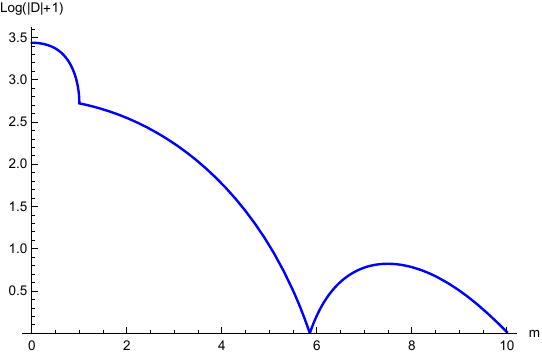}
        \label{fig:left-Dirichlet log plot both mass spectrum}
    \end{subfigure}
    \hfill
    \begin{subfigure}[t]{0.48\textwidth}
        \centering
        \includegraphics[width=\textwidth]{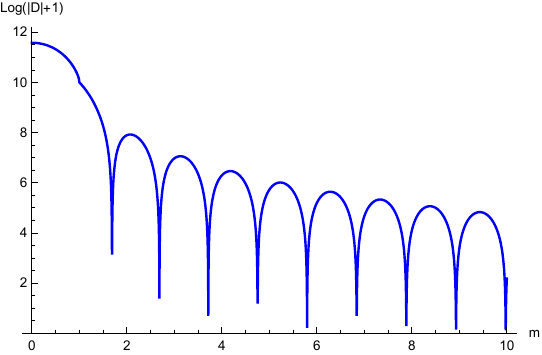}
        \label{fig:right-Dirichlet log plot both mass spectrum}
    \end{subfigure}
    \caption{Plots of $\log|{D(m,\rho_{1},\rho_{2})}+1|$ for $\rho_{1}=1$ and $\rho_{2}=5$ as a function of real $m$ (left) and plots for $\rho_{1}=0.1$ and $\rho_{2}=10$ (right).The downward pointing spike of the graph indicates the zero point of $D$.}
    \label{Dirichlet log plot both mass spectrum}
\end{figure}
\begin{figure}[t]
    \centering
    \begin{subfigure}[t]{0.48\textwidth}
        \centering
        \includegraphics[width=\textwidth]{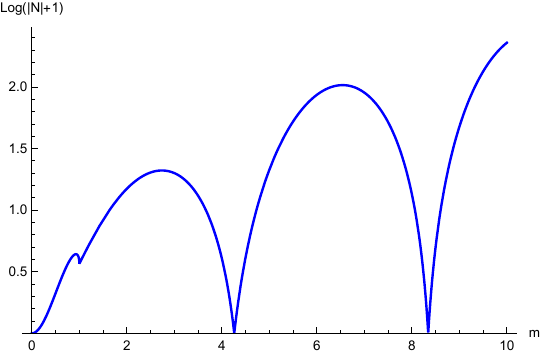}
        \label{fig:left-Neumann log plot both mass spectrum}
    \end{subfigure}
    \hfill
    \begin{subfigure}[t]{0.48\textwidth}
        \centering
        \includegraphics[width=\textwidth]{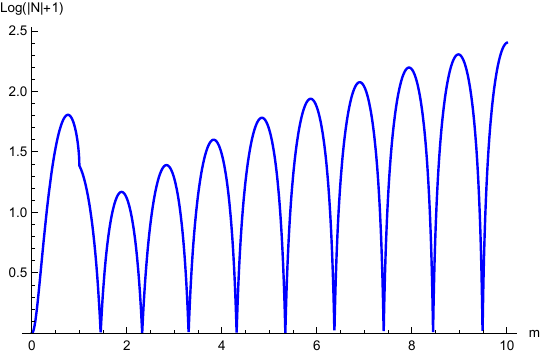}
        \label{fig:right-Neumann log plot both mass spectrum}
    \end{subfigure}
    \caption{Plots of $\log|{N(m,\rho_{1},\rho_{2})}+1|$ for $\rho_{1}=1$ and $\rho_{2}=5$ as a function of real $m$ (left) and plots for $\rho_{1}=0.1$ and $\rho_{2}=10$ (right).The downward pointing spike of the graph indicates the zero point of $N$.}
    \label{Neumann log plot both mass spectrum}
\end{figure}
For better visualization of the spectrum, instead of plotting the functions directly, we consider the logarithmic quantities
\begin{equation*}
\log \left|D^{\mathrm{dS}} \left(m,\rho_1,\rho_2 \right)\right|,
\quad
\log \left|N^{\mathrm{dS}} \left(m,\rho_1,\rho_2 \right) \right|.
\end{equation*}
They are displayed in Figure \ref{Dirichlet log plot both mass spectrum}--\ref{Neumann log plot both mass spectrum}. The downward-pointing spikes appearing in the plots correspond to the zeros of the corresponding spectral functions and therefore determine the allowed KK mass eigenvalues. For the Neumann boundary condition, we find a massless mode $m=0$ in addition to a tower of massive modes. The presence of the zero mode suggests the localization of gravity on the brane, in contrast to the Dirichlet case where no massless mode is present.

The presence of a massless mode with Neumann boundary conditions on the branes can also be determined analytically, see Appendix \ref{appendix:analytical_graviton_ms}. In Appendix \ref{appendix:vector_spectrum}, we have included an analysis of the mass spectrum of a vector field on the ETW brane.

\subsection*{The Volcano Potential}

We can further analyze the localization of the massless graviton mode on the brane by writing the graviton fluctuation equation \eqref{radial eqn} in a Schr\"{o}dinger-like form. This is achieved by considering the coordinate and field redefinitions \cite{Randall:1999vf, Karch:2000ct}
\begin{equation}
    dw=\frac{d\rho}{\sinh(\rho)},\quad \Psi(w)=\sinh^{\frac{d-1}{2}}(\rho)H(\rho).
\end{equation}
With the help of these redefinitions, we rewrite equation \eqref{radial eqn} in the form of the time-independent Schr\"{o}dinger equation
\begin{equation}
    -\Psi''(w) + V(w)\Psi(w) = m^2_{g}\Psi(w),
\end{equation}
where the potential $V(w)$ is given by 
\begin{align}
    \begin{split}
        V(w) &= \frac{d-1}{4} \left( \left(d + 1 \right)\,\mathrm{csch}^{2}(w) + d-1 \right)\\ & \hspace{3 em} + \left(d - 1 \right) \left(\cosh \left(\rho_1 \right)\,\delta \left(w - w_1 \right) - \cosh \left(\rho_2 \right)\,\delta \left(w - w_2 \right) \right).
    \end{split}
\end{align}
This is called a \textit{volcano potential} because of its characteristic shape. Here, \[w_i = \ln \left|\left(\tanh\frac{\rho_i}{2}\right) \right|,\quad i=1,2,\] such that the two branes are located at $w=w_1$ and $w=w_2$, corresponding to $\rho=\rho_1$ and $\rho=\rho_2$, respectively. The delta-function terms arise from the discontinuity of the first derivative of the warp factor across the $Z_2$-orbifold branes. After transforming to the Schrödinger coordinate $w$, these discontinuities appear as localized delta-function contributions to the effective potential.

\begin{figure}[t]
    \centering
    \begin{subfigure}[t]{0.48\textwidth}
        \centering
        \includegraphics[width=\textwidth]{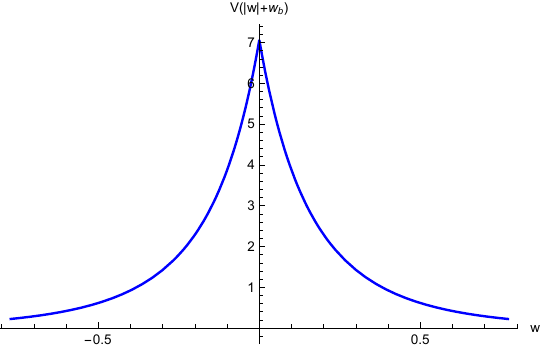}
        \label{fig:left-potential and wavefn plot}
    \end{subfigure}
    \hfill
    \begin{subfigure}[t]{0.48\textwidth}
        \centering
        \includegraphics[width=\textwidth]{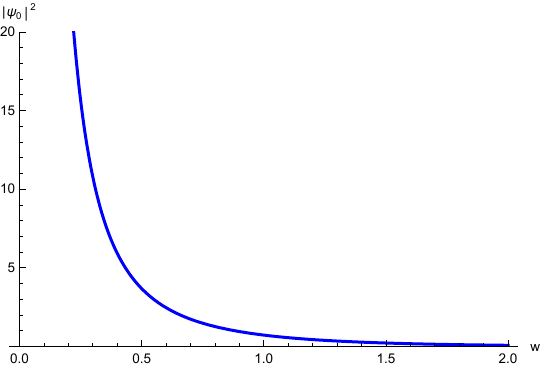}
        \label{fig:right-potential and wavefn plot}
    \end{subfigure}
    \caption{Left: The ${Z}_2$-symmetric volcano potential for ds wedge holography. The delta function potential on the brane is ignored here. Right: Massless graviton wavefunction profile $|\Psi_0(w)|^2$ with the branes located at $w_2 \simeq 0.0135$ ($\rho_2=5$) and $w_1 \simeq 0.7719$ ($\rho_1=2$). we have used $d=3$, and $w_{b}=0.15$ for the above plots.}
    \label{potential and wavefn plot}
\end{figure}

Due to the volcano potential, the massless graviton mode tends to get localized on the branes at low energies. As shown in the right panel of Figure \ref{potential and wavefn plot}, the massless graviton zero-mode wavefunction $\Psi\left(w \right)$ is concentrated near the outer brane $Q_2$, and decays toward the inner brane $Q_1$. This indicates that the zero mode has a larger support on $Q_2$, although it remains non-vanishing throughout the wedge region.

\section{Holographic entanglement entropy} \label{sec:HEE}

A powerful diagnostic tool about the non-local structure of quantum correlations in a bipartite state of a quantum field theory is the entanglement entropy. It quantifies the amount of quantum entanglement shared between a subsystem and its complement. In the context of the AdS/CFT correspondence, the entanglement entropy between a subregion $A$ and its complement $\bar{A}$ in the dual field theory is famously obtained from the bulk geometry by the Ryu-Takayanagi formula \cite{Ryu:2006bv, Ryu:2006ef, Hubeny:2007xt}
\begin{equation} \label{RT formula}
    S_{A}=\frac{\text{Area} \left(\Gamma_{A} \right)}{4G_{N}},
\end{equation}
where $\Gamma_{A}$ is the extremal surface which is homologous to $A$, and attaches to its boundary, i.e., $\partial\Gamma_{A}=\partial A$.

We consider the metric in the wedge geometry $W_{d+1}$
\begin{equation} \label{eq:metric_wedge_for_HEE}
    ds^{2} = d\rho^2 + \sinh^{2}\rho \underbrace{\left(-d\tau^{2} + \cosh^{2}\tau \left(d\theta^2 + \cos^{2}\theta\, d\Omega_{d-2}^{2} \right) \right)}_{\text{dS$_{d}$ slice}},
\end{equation}
with ETW branes located at $\rho = \rho_{1}$ and $\rho = \rho_{2}$. The conformal boundary is reached in the limit $\rho \to \infty$. We always use a large $\rho$ cutoff $\rho_{\infty}$ to facilitate the computation.

We choose a spherically symmetric subregion of angular width $2\theta_{0}$ on the de Sitter future boundary $\tau \rightarrow \infty$. Exploiting the warped character of the background metric \eqref{eq:metric_wedge_for_HEE}, the RT hypersurface can be constructed from a family of extremal hypersurfaces in the dS$_{d}$ slice of bulk AdS$_{d+1}$. We only present the final answer here and relegate the entire computation to Appendix \ref{appendix:RT_hypersurface_dS}.

For $d=3$, i.e., a four-dimensional wedge, we obtain that the holographic entanglement entropy is given by
\begin{equation} \label{eq:HEE_gs_3d_wedge}
    S_{A} = \frac{i \left(\cosh\rho_{2} - \cosh\rho_{1} \right)}{4 G_{N}} \log \left(\frac{\cos^2\theta_{0}}{\epsilon^2}\right),
\end{equation}
where we used equation \eqref{geometric cutoff}. We can compare equation \eqref{eq:HEE_gs_3d_wedge} with the standard result for a CFT$_{2}$ on $S^2$ with a central charge $c$, given by \footnote{
We use the angular convention
$ds^2=d\theta^2+\cos^2\theta\,d\Omega_{d-2}^2$, rather than
$ds^2=d\theta^2+\sin^2\theta\,d\Omega_{d-2}^2$ adopted in \cite{Holzhey:1994we, Calabrese:2004eu, Ogawa:2022fhy}. The two are related by a rotation of the polar axis,
$\theta_{\rm std}=\dfrac{\pi}{2} - \theta$, so that $\sin\theta_0$ appearing in those
references is replaced by $\cos\theta_0$ throughout this work, with no change in the underlying physics.} \cite{Holzhey:1994we,Calabrese:2004eu}
\begin{equation}
    S_{A} = \frac{c}{6}\log\left(\frac{\cos^2 \theta_{0}}{\epsilon^2}\right).
\end{equation}
The comparison tells us that the central charge of the CFT$_{2}$ dual to the wedge $W_{4}$ takes the imaginary value
\begin{equation}
    c = \frac{3 i}{2G_{N}} \left(\cosh\rho_{2} - \cosh\rho_{1} \right),
\end{equation}
which agrees with the value \eqref{central cahrge} inferred from the partition function.\\

\noindent Similarly, for $d=5$, after setting $\tau_{\infty} = -\log\epsilon$ according to \eqref{geometric cutoff}, we obtain
\begin{equation}
    S_{A} = \frac{i \pi\left( \left(\cosh 3\rho_{2} - \cosh 3\rho_{1} \right) - 9 \left(\cosh\rho_{2} - \cosh\rho_{1} \right) \right)}{48 G_{N}} \left(\frac{\cos^2\theta_{0}}{2\epsilon^2} + 2 \log \left(\frac{\cos\theta_{0}}{\epsilon} \right) + \mathcal{O}(1) \right).
\end{equation}
Comparing this with the entanglement entropy of four dimensional CFT \cite{Ryu:2006ef, Hung:2011xb}
\begin{equation}
    S_{A}=c_{0}\,\epsilon^{-2} + 4 a \log \epsilon + \mathcal{O}(1),
\end{equation}
we can read off the value of the central charge $a$
\begin{equation}
     a = -\frac{i\pi}{96G_{N}} \left( \left(\cosh 3\rho_{2} -\cosh 3\rho_{1} \right) - 9 \left(\cosh\rho_{2} - \cosh\rho_{1} \right) \right).
\end{equation}
Indeed, this reproduces our previous estimation \eqref{4d CFT central cgarhe} from the partition function.

\subsection{First law of entanglement entropy}

In this section, we prove that the first law of entanglement entropy is satisfied for dS wedge holography. We focus on NBC because the holographic formula of entanglement entropy is still unknown for wedge holography with DBC. For simplicity, we focus on the first order perturbation including massless and massive graviton fluctuations. Interestingly, we find that the massive modes do not change the holographic entanglement entropy.\\

To begin, let us give a brief review of the first law of entanglement. Given two density matrices $\rho_0$ and $\rho_1$, their relative entropy is
\begin{equation}
    S \left(\rho_1|\rho_0 \right) = \mathrm{Tr} \left(\rho_1\log\rho_1 \right) - \mathrm{Tr} \left(\rho_1\log\rho_0 \right) .
\end{equation}
In general, relative entropy is positive \cite{Wehrl:1978zz, Vedral:2002zz, Blanco:2013joa}
\begin{equation}
    S \left(\rho_1|\rho_0 \right) \geq 0,
\end{equation}
and vanishes only when the two states are identical. Introducing the modular Hamiltonian $H_{0}$ associated with the reference state $\rho_0$
\begin{equation}
    \rho_0 = \frac{e^{-H_0}}{{\rm Tr}\,e^{-H_0}},
\end{equation}
one may rewrite the relative entropy as
\begin{equation}
    S \left(\rho_1|\rho_0 \right) = \Delta \langle H_0 \rangle - \Delta S,
\end{equation}
where
\begin{equation}
    \Delta \langle H_0\rangle = {\rm Tr} \left(\rho_1 H_0 \right) - {\rm Tr} \left(\rho_0 H_0 \right), \quad
    \Delta S = S \left(\rho_1 \right) - S \left(\rho_0 \right),
\end{equation}
and $S \left(\rho \right) = -{\rm Tr} \left(\rho\log\rho \right)$ is the entanglement entropy. Positivity of relative entropy, therefore, implies
\begin{equation}
    \Delta \langle H_0\rangle \geq \Delta S .
\end{equation}
If $\rho_1$ is an infinitesimal perturbation of the reference state, \[\rho_{1} \left(\epsilon \right) = \rho_0 + \epsilon\, \delta\rho,\] the first-order variation of relative entropy vanishes\footnote{This follows from $S\left(\rho_{1}(0)| \rho_{0} \right) = 0$, and $S\left(\rho_{1}\left(\epsilon \right)| \rho_{0} \right) > 0$ for both positive and negative $\epsilon$.}, giving
\begin{equation}
    \Delta S = \Delta \langle H_0\rangle .
\label{first-law-entanglement}
\end{equation}
This relation is known as the first law of entanglement \cite{Blanco:2013joa}.

For a general region the modular Hamiltonian is non-local and difficult to write explicitly. However, for the vacuum state of a CFT and a spherical subregion $B$ of radius $\ell$, it is local and is known to be given by \cite{Casini:2011kv, Wong:2013gua}
\begin{equation}
    H_B = 2\pi \int_{r<\ell} d^{d-1}x \frac{\ell^2-r^2}{2\ell} T_{00}(x),
\end{equation}
where $T_{00}$ is the time-time component of the CFT stress-energy tensor. For homogeneous excitations this reduces the entanglement first law to a form closely analogous to the ordinary first law of thermodynamics
\begin{equation} \label{eq:first_law_of_HEE}
    T_E\, \Delta S_B = \Delta E_B ,
\end{equation}
where $\Delta E_B$ is the change of energy inside the entangling region and
\begin{equation}
    T_E \sim \frac{1}{\ell}
\end{equation}
is called the \emph{entanglement temperature}. Generally, the numerical coefficient in $T_E \sim 1/\ell$ depends on the shape and size $\ell$ of the subsystem, and the dimensionality of the spacetime; but the inverse-size scaling is universal for small subsystems in relativistic theories. This thermodynamic interpretation of entanglement entropy was emphasized holographically in \cite{Bhattacharya:2012mi, Allahbakhshi:2013rda}.

When the theory contains additional conserved charges, the modular Hamiltonian may also contain the corresponding charge operators. In such cases the first law is modified by chemical-potential-like terms. Schematically, one obtains
\begin{equation} 
    T_E\, \Delta S_B = \Delta E_B - \mu_E\,\Delta Q_B.
\end{equation}
Here $\mu_E$ is an \emph{entanglement chemical potential} conjugate to the conserved charge $Q_B$ \cite{Wong:2013gua}.

In holographic CFTs, the first law of entanglement has an important implication. For spherical subregions, imposing the entanglement first law for any state infinitesimaly close to the CFT vacuum is equivalent to the linearized Einstein's equations of motion in the dual asymptotically AdS spacetime \cite{Lashkari:2013koa, Faulkner:2013ica, Faulkner:2017tkh, Paul:2018spp}.

\subsubsection{The massive fluctuations}

We first discuss the massive graviton fluctuations. Consider the metric ansatz \eqref{perturbed metric}
\begin{equation}\label{metric for massive flu}
    ds^2=d\rho^2+\sinh^2\rho\left(\frac{-dz^2+\delta_{ij} dx^idx^j}{z^2}+\epsilon\sum_{m}H^m(\rho)h_{ij}(x)dx^idx^j\right),
\end{equation}
where $x^\mu=(z,x^i), x^i=(t_E,x^A)=(t_E,x^1\ldots,x^{d-2})$, and $\sum_{m}$ denotes a sum over the mass spectrum. 
Here $t_E$ denotes the Euclidean time coordinate on the boundary. Throughout this work, we consider the entangling region on the constant Euclidean time slice $t_E=0$. We are interested in the holographic entanglement entropy associated with a $(d-2)$-dimensional spherical subsystem \[\sum_{A=1}^{d-2}(x^A)^2\le \ell^2,\quad x_{d-1} = 0.\]
on the Euclidean boundary at fixed Euclidean time $t_E = x_{d-1} = 0.$ For a background metric \eqref{metric for massive flu} with $\epsilon=0$, the Corresponding RT hypersurface in the bulk is given by \cite{Ogawa:2022fhy, Akal:2020wfl}
\begin{equation}\label{eq:RT in first law}
    x^2-z^2=\ell^2, 
\end{equation}
where $x^2=\sum_{A=1}^{d-2}(x^A)^2$. At the leading order $O\left(\epsilon \right)$, the HEE can be computed by evaluating the area of the zeroth order extremal hypersurface \eqref{eq:RT in first law} in the perturbed geometry \eqref{metric for massive flu}. The induced metric on the RT surface is given by
\begin{equation}
    ds^2_{\gamma}=d\rho^2+\frac{\sinh^2\rho}{z^2}\gamma_{AB}dx^Adx^B,
\end{equation}
where 
\begin{equation}
    \gamma_{AB}=\delta_{AB}-\partial_{A}
z\partial_{B} z+\epsilon z^2\sum_{m}H^{m}(\rho)h^{(m)}_{AB}(x); 
\end{equation}
By applying the RT formula eq \eqref{RT formula}, we get the holographic entanglement entropy
\begin{equation}
    S=\frac{1}{4G_{N}}\int_{\rho_{1}}^{\rho_{2}} d\rho\int_{x^2\leq\ell^2}d^{d-2}x \frac{\sinh^{d-2}\rho}{z^{d-2}}\sqrt{|\gamma_{AB}|}.
\end{equation}
Therefore, the first order variation of entanglement entropy is easily found to be
\begin{equation} \label{eq:change in entropy}
    \delta S=\sum_{m}\frac{\epsilon}{8G_{N}}\int_{\rho_{1}}^{\rho_{2}} d\rho  \ \sinh^{d-2}\rho  \ H^{m}(\rho)\int_{x^2\leq\ell^2}d^{d-2}x f^{(m)}(x^A),
\end{equation}
with 
\begin{equation}
    f^{(m)}(x^A)=\frac{1}{z^{d-4}}\left(\delta^{CD}h_{CD}\sqrt{1-\delta^{AB}\partial_{A}z\partial_{B}z}+\frac{h^{CD}\partial_{C}z\partial_Dz}{\sqrt{1-\delta^{AB}\partial_{A}z\partial_{B}z}}\right).
\end{equation}
Here $h^{CD}$ are raised by $\delta^{CD}$ and $z(x)$ denotes the ground state RT surface in Equation \eqref{eq:RT in first law}. In this derivations, we have used the formula \cite{Blanco:2013joa}
\begin{equation*}
    \left| g_{AB} - \partial_{A} z\,\partial_{B} z \right| = \left|g_{AB} \right| \left(1 - g^{AB} \partial_{A}z \partial_{B}z \right).
\end{equation*}

To proceed further, we establish the orthogonality relation satisfied by the radial wavefunctions $H^{(n)}\left(\rho\right)$. The radial equation \eqref{radial eqn} can be rewritten as
\begin{equation}
    \frac{d}{d\rho} \left(\sinh^{d}\rho\, \frac{d H^{(m)} \left(\rho \right)}{d\rho} \right) + m^{2} \sinh^{d-2}\rho\, H^{(m)} \left(\rho \right) = 0.
    \label{SL_form}
\end{equation}
This is a Sturm--Liouville equation with weight function
\begin{equation}
    w(\rho)=\sinh^{d-2}\rho .
\end{equation}
Let $H^{(m)}(\rho)$ and $H^{(n)}(\rho)$ be two eigenfunctions corresponding to eigenvalues $m^{2}$ and $n^{2}$. From the general equation \eqref{SL_form} we obtain
\begin{equation}
\frac{d}{d\rho}
\left[
\sinh^{d}\rho
\left(
H^{(n)}\frac{dH^{(m)}}{d\rho}
-
H^{(m)}\frac{dH^{(n)}}{d\rho}
\right)
\right]
+
(m^{2}-n^{2})
\sinh^{d-2}\rho\,
H^{(m)}H^{(n)}
=
0.
\end{equation}
Integrating between the two branes located at $\rho=\rho_{1}$ and $\rho=\rho_{2}$
\begin{equation}
    \left(m^{2} - n^{2} \right) \int_{\rho_{1}}^{\rho_{2}} d\rho\, \sinh^{d-2}\rho\,H^{(m)}H^{(n)} = - \left[\sinh^{d}\rho \left(H^{(n)}\frac{dH^{(m)}}{d\rho} - H^{(m)} \frac{dH^{(n)}}{d\rho} \right) \right]_{\rho_{1}}^{\rho_{2}}.
\end{equation}
The boundary term vanishes for both Dirichlet and Neumann boundary conditions. Therefore, for $m^{2}\neq n^{2}$
\begin{equation}
\int_{\rho_{1}}^{\rho_{2}}
d\rho\,
\sinh^{d-2}\rho\,
H^{(m)}(\rho)\,
H^{(n)}(\rho)
=
0.
\end{equation}
We can combine this with the normalization condition to write the orthonormality relation for radial graviton wavefunction
\begin{equation}
    \int_{\rho_{1}}^{\rho_{2}} d\rho\,\sinh^{d-2}\rho\, H^{(m)}(\rho)\, H^{(n)}(\rho) = N_m\,\delta_{mn},
    \label{orthogonality}
\end{equation}
where $N_m$ is a normalization constant.

For the massless mode, $m=0$, equation \eqref{SL_form} reduces to
\begin{equation}
    \sinh^{d}\rho \frac{dH^{(0)}}{d\rho} = C,
\end{equation}
where $C$ is a constant. Imposing Neumann boundary conditions on both branes, we get $C=0$, and hence
\begin{equation}
    \frac{dH^{(0)}}{d\rho}=0.
\end{equation}
Therefore, the massless radial mode is simply
\begin{equation}
    H^{(0)}(\rho)=\text{constant}.
\label{constant_zero_mode}
\end{equation}

Substituting equation\eqref{constant_zero_mode} into equation\eqref{orthogonality} with $n=0$, we obtain
\begin{equation}
    \int_{\rho_{1}}^{\rho_{2}} d\rho\, \sinh^{d-2}\rho\, H^{(m)}(\rho) = 0,\quad m>0.
    \label{vanishing_integral}
\end{equation}
This is exactly the radial integral in equation \eqref{eq:change in entropy}. As a consequence, the first-order correction to the entanglement entropy \eqref{eq:change in entropy} vanishes for all massive fluctuations
\begin{equation}
    \delta S = 0.
\end{equation}

The first-order variation of the expectation value of the modular Hamiltonian associated with the Euclidean spherical entangling region is given by
\begin{equation}\label{eq:modular Hamiltonian}
\delta\langle H\rangle
=
\frac{\pi}{\ell}
\int_{x\le\ell}
d^{d-2}x\,
z^2\,
\delta\langle T_{t_Et_E}\rangle,
\end{equation}
where $T_{t_Et_E}$ denotes the Euclidean stress tensor component along the Euclidean time direction. Using the RT surface \eqref{eq:RT in first law},
\begin{equation}
    z^2=x^2-\ell^2,
\end{equation}
Equation \eqref{eq:modular Hamiltonian} can be rewritten as
\begin{equation}
    \delta\langle H\rangle = \frac{\pi}{\ell} \int_{x\le\ell} d^{d-2}x\, (x^2-\ell^2)\, \delta\langle T_{t_Et_E}\rangle = -\frac{\pi}{\ell} \int_{x\le\ell} d^{d-2}x\, (\ell^2-x^2)\, \delta\langle T_{t_Et_E}\rangle.
\end{equation}
To evaluate the stress-tensor holographically, we apply the dS$_{d}$/CFT$_{d-1}$ correspondence on the brane \cite{Strominger:2001pn}. As shown in Appendix \ref{appendix:enmomtensor}, the holographic stress tensor is obtained by analytically continuing the AdS/CFT result, and it is given by
\begin{equation}
    \langle T_{ij} \rangle=i\frac{(d-1)}{16\pi G^{(d)}_{N}}h^{(d-1)}_{ij},
\end{equation}
where $G^{(d)}_{N}$ is the effective Newton's constant on the brane, and $h^{(d-1)}_{ij}$ is the coefficient in the near-boundary Fefferman-Graham (FG) expansion of the induced metric on the brane 
\begin{equation}\label{FG expansion}
    ds_Q^2=\frac{-dz^2+\left(\delta_{ij}+z^{d-1}h^{(d-1)}_{ij}+...\right)dx^idx^j}{z^2}.
\end{equation}
For massive KK graviton modes, the FG expansion takes the form \cite{Hu:2022lxl}
\begin{equation}\label{FG expansion massive mode}
    ds_Q^2=\frac{-dz^2+\left(z^{d-1-\Delta}h^{(d-1-\Delta)}_{ij}+z^{\Delta}h^{(\Delta)}_{ij}...\right)dx^idx^j}{z^2},
\end{equation}
where \[\Delta=\frac{d-1}{2}+\sqrt{\frac{(d-1)^2}{4}-m^2}\] is the conformal dimension of the massive graviton.

Comparing equation \eqref{FG expansion} with \eqref{FG expansion massive mode} , we get $\langle T_{ij} \rangle\sim h^{d-1}_{ij}=0$ for the massive modes. As a result the first order variation of the expectation value of the modular Hamiltonian in equation \eqref{eq:modular Hamiltonian} also vanishes. Thus we verify the first law of entanglement entropy for the massive fluctuations
\begin{equation}
    \delta S=\delta \langle H \rangle=0,\ \text{for} \ m^2>0.
\end{equation}

\subsubsection{The massless fluctuations}

Next, we consider the massless graviton fluctuations. In this case, the $m=0$ mode in Equation \eqref{metric for massive flu} is given by
\begin{equation}\label{eq:massless fluctuation}
    h^{m=0}_{ij}(x^\mu)
    =
    z^{d-3}h^{d-1}_{ij}(x^i)
    =
    -i\frac{16\pi G_{N}^{(d)}}{d-1}
    z^{d-3}
    \langle T_{ij}(x^i)\rangle,
\end{equation}
where the effective Newton's constant is
\begin{equation}
    \frac{1}{G_{N}^{(d)}}
    =
    \frac{1}{G_{N}}
    \int_{\rho_{1}}^{\rho_{2}}
    d\rho\,
    \sinh^{d-2}\rho.
\end{equation}
For convenience, we normalize the radial wavefunction of the massless mode by choosing
\[
H^{(m=0)}(\rho)=1.
\]
Substituting Equations \eqref{eq:RT in first law} and \eqref{eq:massless fluctuation} into the linearized entropy variation \eqref{eq:change in entropy}, we obtain the first-order correction to the holographic entanglement entropy,
\begin{equation}\label{eq:delta s for massless mode}
    \delta S
    =
    \frac{2\pi \ell \Omega_{d-4}}{d-1}
    \int_{0}^{\ell}
    dx\,x^{d-3}
    \int_{0}^{\pi}
    d\theta\,
    \sin^{d-4}\theta
    \left(
    \langle T_{~A}^{A}(x)\rangle
    -
    \langle T_{AB}(x)\rangle
    \frac{x^Ax^B}{\ell^2}
    \right),
\end{equation}
where $\Omega_{d-4}$ denotes the volume of the unit $(d-4)$-sphere.

Similarly, using Equation \eqref{eq:modular Hamiltonian}, the first-order variation of the modular Hamiltonian takes the form
\begin{equation}\label{eq:modular Hamiltonian for massless mode}
    \delta\langle H\rangle
    =
    \frac{\pi \Omega_{d-4}}{\ell}
    \int_{0}^{\ell}
    dx\,x^{d-3}
    \int_{0}^{\pi}
    d\theta\,
    \sin^{d-4}\theta
    \left(x^2-\ell^2\right)
    \langle T_{t_Et_E}(x)\rangle.
\end{equation}

To facilitate the comparison between Equations \eqref{eq:delta s for massless mode} and \eqref{eq:modular Hamiltonian for massless mode}, we express the boundary stress tensor in momentum space through the Fourier expansion
\begin{equation}\label{eq:Fourier expansion}
    T_{ij}(x) = \int d^{\,d-1}p\, e^{-ip\cdot x}\, T_{ij}(p).
\end{equation}
Following \cite{Blanco:2013joa}, we choose the spatial momentum to lie along the $x^1$-direction. The conservation and tracelessness of the Euclidean stress tensor imply the identities
\begin{equation} \label{eq:ward identity}
\langle \delta^{A B}\, T_{A B}\rangle = -\langle T_{t_Et_E}\rangle,\quad \langle T_{1\, t_E}\rangle = -\frac{p^{t_E}}{p^1}\langle T_{t_Et_E}\rangle,\quad \langle T_{11}\rangle = \left(\frac{p^{t_E}}{p^1}\right)^2 \langle T_{t_Et_E}\rangle.
\end{equation}
In the first equation above, $A, B = 1, 2, \ldots, d-2$. Next, we evaluate the integral appearing in Equation \eqref{eq:delta s for massless mode}. Since the integrand in Equation \eqref{eq:delta s for massless mode} is rotationally symmetric about the $x^1$-direction, all off-diagonal contributions proportional to $T_{AB}x^Ax^B$ with $A\neq B$ vanish upon integration. Furthermore, by rotational symmetry, the diagonal contributions corresponding to $A=B=2,\ldots,d-2$ are identical. Consequently, the integrand can be rewritten as
\begin{equation}
    \begin{split}
        \langle T_{~A}^{A}(p)\rangle - \langle T_{AB}(p)\rangle \frac{x^Ax^B}{\ell^2} &= - \langle T_{t_Et_E}\rangle - \langle T_{11}\rangle \frac{(x^1)^2}{\ell^2} - \sum_{A=2}^{d-2} \langle T_{AA}\rangle \frac{(x^A)^2}{\ell^2} \\
        &= - \langle T_{t_Et_E}\rangle - \langle T_{11}\rangle \frac{(x^1)^2}{\ell^2} - \sum_{A=2}^{d-2} \langle T_{AA}\rangle \frac{x^2-(x^1)^2}{(d-3)\ell^2} \\
        &= \langle T_{t_Et_E}\rangle \left(- 1 - \left(\frac{p^{t_E}}{p^1}\right)^2 \frac{(x^1)^2}{\ell^2} + \frac{1 + \left(\frac{p^{t_E}}{p^1}\right)^2}{d-3} \frac{x^2-(x^1)^2}{\ell^2} \right).
    \end{split}
\end{equation}
Finally, in the small-momentum limit, $p\to 0$, we obtain
\begin{equation} \label{identity 1}
    \langle T_{~A}^{A}(p)\rangle - \langle T_{AB}(p)\rangle \frac{x^Ax^B}{\ell^2} \simeq -\langle T_{t_Et_E}\rangle \left(\frac{\ell^2-x^2\cos^2\theta}{\ell^2} + \mathcal{O}\left(p^2 \right) \right),
\end{equation}
where
\[p^{t_E} = \sqrt{p^2 - \left(p^1 \right)^2}. \]
Finally, the variation of the holographic entanglement entropy can be expressed in polar coordinates as
\begin{equation}\label{eq:delta s in polar form}
    \delta S = -\frac{2\pi \ell\, S_{d-4}}{d-1}\langle T_{t_E t_E}\rangle\int_{0}^{\ell} dx\,x^{d-3}\int_{0}^{\pi} d\theta\,\sin^{d-4}\theta\,e^{-ip^1x\cos\theta}\left(\frac{\ell^2-x^2\cos^2\theta}{\ell^2}\right).
\end{equation}
The angular integration can be carried out using the identity
\begin{equation} \label{eq:int_identity-1}
\int_{0}^{\pi}d\theta\,\sin^{m}\theta\,e^{-ix\cos\theta}
=
2^{m/2}\sqrt{\pi}\,
\Gamma\!\left(\frac{m+1}{2}\right)
\frac{J_{m/2}(|x|)}{|x|^{m/2}}.
\end{equation}
The remaining radial integral is evaluated with the standard Bessel function identity
\begin{equation} \label{eq:int_identity-2}
\int_{0}^{\ell}dx\,x^{\nu+1}J_{\nu}(ax)
=
\frac{\ell^{\nu+1}}{a}J_{\nu+1}(a\ell),
\end{equation}
where $J_{\nu}$ denotes the Bessel function of the first kind.

On the other hand, from Equation \eqref{eq:modular Hamiltonian for massless mode}, the first-order variation of the modular Hamiltonian is
\begin{equation}
    \delta\langle H\rangle = - \frac{\pi\, \Omega_{d-4}}{\ell} \langle T_{t_Et_E}\rangle \int_{0}^{\ell}dx\,x^{d-3}     \int_{0}^{\pi}d\theta\, \sin^{d-4}\theta\,e^{-ip^1x\cos\theta} \left(\ell^2 - x^2 \right).
\end{equation}
Using the integral identities \eqref{eq:int_identity-1}-\eqref{eq:int_identity-2}, we obtain
\begin{equation}\label{eq:delta H result}
    \delta\langle H\rangle
    =
    -2^{\frac{d-2}{2}}
    \pi^{3/2}
    \Omega_{d-4}
    \Gamma\!\left(\frac{d-3}{2}\right)
    \frac{\ell^{\frac{d-2}{2}}}{|p^1|^{\frac{d}{2}}}
    J_{\frac{d}{2}}\!\left(|p^1|\ell\right).
\end{equation}

Similarly, evaluating the integrals in Equation \eqref{eq:delta s in polar form} yields
\begin{equation}
    \delta S
    =
    -\frac{2\pi \ell\, \Omega_{d-4}}{d-1}
    \langle T_{t_Et_E}\rangle
    \int_{0}^{\ell}dx\,x^{d-3}
    \int_{0}^{\pi}d\theta\,
    \sin^{d-4}\theta\,
    e^{-ip^1x\cos\theta}
    \left(\frac{\ell^2-x^2\cos^2\theta}{\ell^2}\right).
\end{equation}
Applying the same integral identities, we arrive at
\begin{equation}\label{eq:delta S result}
    \delta S
    =
    -2^{\frac{d-2}{2}}
    \pi^{3/2}
    \Omega_{d-4}
    \Gamma\!\left(\frac{d-3}{2}\right)
    \frac{\ell^{\frac{d-2}{2}}}{|p^1|^{\frac{d}{2}}}
    J_{\frac{d}{2}}\!\left(|p^1|\ell\right),
\end{equation}
where $J_{\nu}$ is the Bessel function of the first kind. Comparing Equations \eqref{eq:delta H result} and \eqref{eq:delta S result}, we obtain
\[
\delta\langle H\rangle=\delta S,
\]
thereby establishing the first law of holographic entanglement entropy in the dS wedge holography framework.

\subsection{Timelike entanglement entropy between two codimension-two defects}

There is an alternative notion of entropy in wedge holographic setting that considers the entanglement between the two defects located where the two branes intercept \cite{Akal:2020wfl, Geng:2020fxl}. In our construction of wedge holography, gravity in the $(d+1)$-dimensional wedge $W_{d+1}$ is dual to a $(d-1)$-dimensional CFT living on codimension-two defect located at the intersection of the branes \cite{Ogawa:2022fhy, Aguilar-Gutierrez:2023tic, Hao:2025ocu}. There are two such defects, represented by a pair of $S^{d-1}$ situated at the tips of the wedge $\left(\tau = \pm \infty \right)$ near $\mathcal{I}^{+}$. Since these defects are timelike separated, the standard notion of entanglement entropy is not applicable. Instead, one may consider the \emph{pseudo entropy} or \emph{timelike entanglement entropy} introduced in \cite{Nishioka:2021cxe, Mollabashi:2021xsd, Chen:2023gnh, Doi:2022iyj, Jiang:2023loq, Doi:2023zaf}. Holographically, this quantity is captured by the area of an extremal hypersurface connecting subregions on the two defects.
\begin{figure}[t]
    \centering 
    \includegraphics[width =0.5\linewidth, height=9cm]{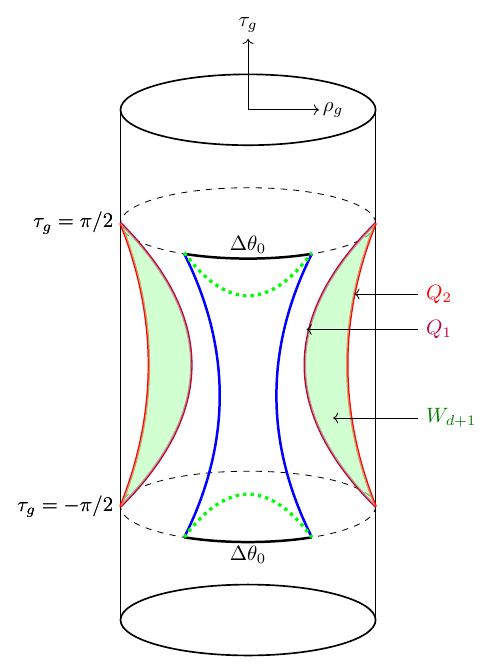} 
    \caption{Candidate extremal hypersurfaces between timelike separated codimension-2 defects. The connected surface that captures timelike entanglement is shown in blue.} 
    \label{Conn phase pic}
\end{figure}

 We consider two rigid branes placed very close to each other. The line element of AdS$_{d+1}$ foliated by $d$-dimensional de Sitter slices is given by
    \begin{equation}
         ds^2 = d\rho^2 + \sinh^2\rho \left(-d\tau^2 + \cosh^2\tau\,  d\Omega_{d-1}^2 \right).
    \end{equation}
 We want to evaluate the area of an extremal hypersurface at  fixed $\rho=\rho_{0}$ -- which extends between $-\theta_{0}$ and $\theta_{0}$ on the boundary sphere $S^{d-1}$, and is anchored on the asymptotic slice $\tau = \pm\tau_{0} \to \infty$. To this end, we need to compute
\begin{equation} \label{Area in timelike}
    A=2\, i\, \Omega_{d-2} \left(\sinh\rho_{0} \right)^{d-1} \int_{-\tau_{0}}^{\tau_{0}} d\tau  \left(\cosh\tau \cos\theta \left(\tau \right) \right)^{d-2} \sqrt{1 - \theta' \left(\tau \right)^2 \cosh^2\tau},
\end{equation}
where $\Omega_{d-2}$ is the volume element of $S^{d-2}$. The overall factor of $i$ in equation\eqref{Area in timelike} originates from the Lorentzian signature of the induced metric on the extremal hypersurface. The factor of 2 arises from $\mathbb{Z}_{2}$ symmetry. See Figure \ref{Conn phase pic} for a schematic representation of the timelike hypersurface in our geometric setup. From equation \eqref{Area in timelike}, we derive the following Euler-Lagrange equation
\begin{align} \label{EOM timelike}
    \begin{split}
        \left(d - 2 \right) \sin\theta &+ \cosh\tau \left(\frac{d^{2}\theta}{d\tau^{2}}\cos\theta \cosh\tau + d \frac{d\theta}{d\tau} \cos\theta \sinh\tau \right.\\ &\left. + \left(d - 2 \right) \left(\frac{d\theta}{d\tau} \right)^{2} \sin\theta \cosh\tau - \left(d - 1 \right) \left(\frac{d\theta}{d\tau} \right)^{3} \cos\theta \cosh^{2}\tau \sinh\tau \right) = 0.
    \end{split}
\end{align}
 We solve the equation of motion \eqref{EOM timelike} numerically by imposing Dirichlet boundary conditions on the profile $\theta(\tau)$. For concreteness, we consider the case $d=4$ and choose an angular interval of width $\theta_{0} - \left(-\theta_{0} \right) = \Delta\theta_{0} = \pi/4$, such that the extremal surface is anchored at $\theta(\pm\tau_{0})=\pm\theta_{0}$. The connected timelike extremal surface satisfies the reflection symmetry $\theta\left(-\tau \right) = -\theta\left(\tau \right)$, which implies $\theta\left(\tau = 0 \right) = 0$.
 
\begin{figure}[t]
    \centering
    \begin{subfigure}{0.48\textwidth}
        \centering
        \includegraphics[width = \linewidth]{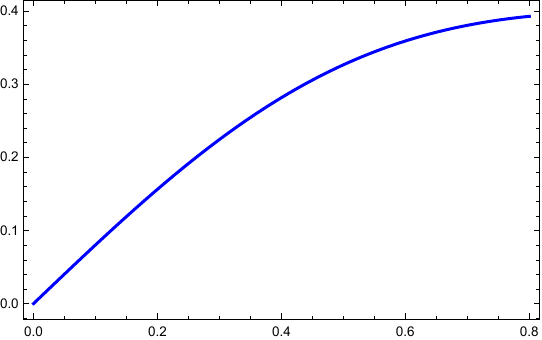}
        \label{fig:timelike_RT_surface_profile}
    \end{subfigure}
    \hfill
    \begin{subfigure}{0.48\textwidth}
        \centering
        \includegraphics[width=\linewidth]{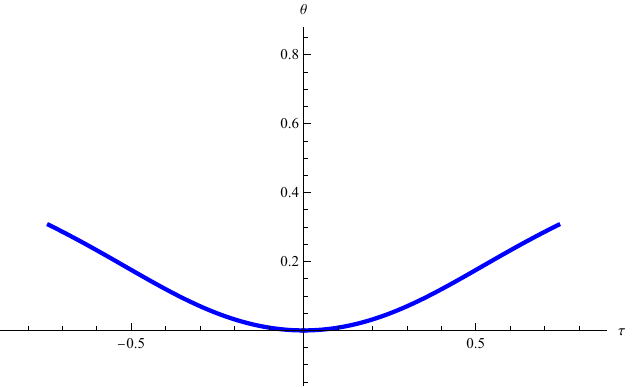}
        \label{fig:timelike_RT_surface_polar_profile}
    \end{subfigure}
    \caption{Connected timelike extremal surface for $d=4$ with $\Delta\theta_{0}=\pi/4$. Left: numerical profile $\theta(\tau)$ obtained by solving the Euler--Lagrange equation with Dirichlet boundary conditions. Right: polar representation of the extremal surface, illustrating the connected timelike RT surface stretching between the two codimension-2 defects.}
    \label{fig:Timelike RT both plot}
\end{figure}

 The numerical profile and its corresponding polar representation are shown in Figure \ref{fig:Timelike RT both plot}.
Substituting this solution into the area functional\eqref{Area in timelike} and performing the integral over the specified range, we obtain the timelike entropy as
\begin{equation}
    S_{\mathrm{con}} = \frac{A}{4G_{N}}= 1.3396\,  i
\end{equation}
In this case, we fixed the brane position at $\rho = \rho_{0} = 0.5$, and set the Newton constant $G_N$ to unity. Therefore, the holographic timelike entropy is \emph{purely imaginary}, which is fully consistent with the expected behavior of pseudo entropy in Lorentzian signature \cite{Aguilar-Gutierrez:2023tic}.

Our result differs from the original pseudo-entropy constructions of \cite{Doi:2022iyj}, where extremal surfaces typically contain Euclidean or mixed-signature segments that contribute to the real part of the pseudo entropy. In the present wedge-holography setup, the extremal surface remains timelike throughout, so the induced metric is Lorentzian everywhere. Consequently, the area functional carries an overall factor of i, yielding a purely imaginary pseudo entropy.

In addition to the connected extremal surface, one may also consider a disconnected configuration consisting of two independent surfaces attached to the individual defects, see the two dotted green curves in Figure \ref{Conn phase pic}. In the computation above, we considered that the two branes were located close to each other. In this limit, the corresponding codimension-2 defects remain strongly correlated, and a connected timelike extremal surface stretching between them is expected to provide the dominant contribution to the pseudo entropy. For sufficiently large brane separations, disconnected extremal surfaces may become competitive, leading to a possible phase transition.

\section{Information recovery in dS wedge holography} \label{sec:Page_curve}

In this section, we investigate a Page-like transition in dS wedge holography for general $d > 2$, extending the analysis of \cite{Aguilar-Gutierrez:2023tic}.

At the outset, the geometry \eqref{dS slicing metric} does not contain a dynamically evaporating black hole. Rather, we consider that the gravitating infrared sector, geometrized by the IR brane $Q_1$ at $\rho=\rho_1$, plays the role of the evaporating black hole. The ultraviolet sector associated with $Q_2$ plays the role of the radiation bath. A subregion on $Q_{2}$ is interpreted as the bath where radiation is collected by an external observer.

To obtain a sharply defined radiation subsystem, we take the UV brane towards the asymptotic AdS boundary,
\begin{equation*}
\rho_2 \rightarrow \infty,
\end{equation*}
so that gravity decouples on $Q_2$. In this limit, extremal surfaces can be anchored on a prescribed region of $Q_2$ with Dirichlet boundary conditions \cite{Geng:2020fxl}, while gravity remains dynamical on the IR brane $Q_1$. The total Hilbert space is assumed to get factorized as
\begin{equation}
\mathcal{H}_{\rm total}
=
\mathcal{H}_{\rm UV}
\otimes
\mathcal{H}_{\rm IR},
\end{equation}
where $\mathcal{H}_{\rm UV}$ is the Hilbert space of an effective dS QFT on $Q_{2}$, and $\mathcal{H}_{\rm IR}$ represents the IR d.o.f. containing both gravity and field theory parts.\\

\noindent We consider a rotationally symmetric radiation region
\[R = I\times S^{d-2},\quad I = \lbrace \left. \theta \right| -\theta_0 \leq \theta \leq \theta_0 \rbrace, \]
on a constant-$\tau$ slice of global de Sitter spacetime
\begin{equation}
    ds^2 = -d\tau^2 + \cosh^2\tau \left(d\theta^2 + \cos^2\theta\, d\Omega_{d-2}^2 \right).
\end{equation}
Accordingly, $\mathcal{H}_{\rm UV} = \mathcal{H}_{R} \otimes \mathcal{H}_{R^c},$ and the radiation entropy is
\begin{equation}
    S_R = -\mathrm{Tr} \left(\rho_R \log \rho_R \right).
\end{equation}
The holographic entanglement entropy is obtained from the generalized HRT prescription \cite{Ryu:2006bv, Ryu:2006ef, Hubeny:2007xt}
\begin{equation} \label{Entropy formula}
    S_{\rm HEE} = \frac{1}{4 G_{N}}\, \min \left\{\mathrm{ext} \left(\int_\Gamma d^{\,d-1}x \sqrt{\gamma} \right) \right\},
\end{equation}
where $\Gamma$ is a codimension-two extremal surface homologous to $R$, and $\gamma$ is its induced metric.

\begin{figure}[t]
    \centering
    \includegraphics[width=0.46\linewidth]{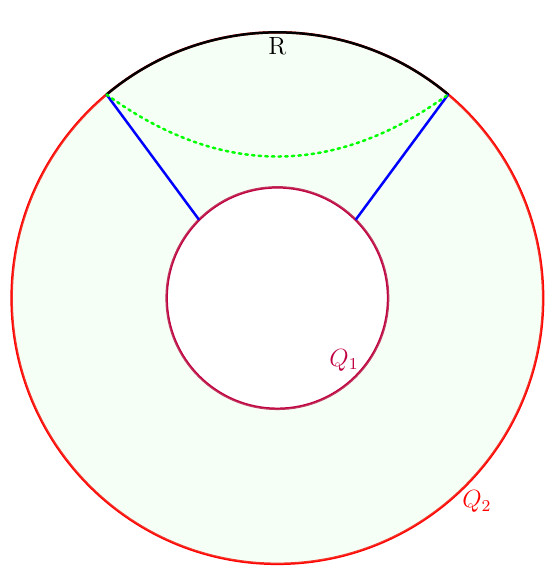}
    \caption{A two-dimensional projection of $W_{d+1}$ with the two classes of competing HRT hypersurface. The green, disconnected surface dominates in the no-island phase, while the blue, connected surface ending on the IR brane dominates after the Page transition.}
    \label{RT on island phase}
\end{figure}

There are two classes of candidate extremal hypersurfaces as illustrated in Figure \ref{RT on island phase}. For sufficiently small radiation regions, the disconnected HRT surface anchored entirely on $Q_{2}$ is expected to dominate; it represents the no-island phase. As the physical size of $R$ increases, a blue connected HRT surface that extends from $Q_{2}$ to $Q_{1}$ becomes dominant, signalling the appearance of an entanglement island and the Page transition.

For sufficiently large subregion, the homology constraint may allow additional saddles involving an additional hypersurface wrapping the minimal cross-section of the IR brane. In this work, we restrict ourselves to $\Delta\theta\le\pi,$ so that the homology transition does not arise. Unlike \cite{Aguilar-Gutierrez:2023tic}, which mainly considers $d=2$, our analysis is valid in arbitrary spacetime dimensions.
%

\subsection{No-island Phase} \label{No-island}

\begin{figure}[t]
    \centering
    \includegraphics[width=0.65\textwidth]{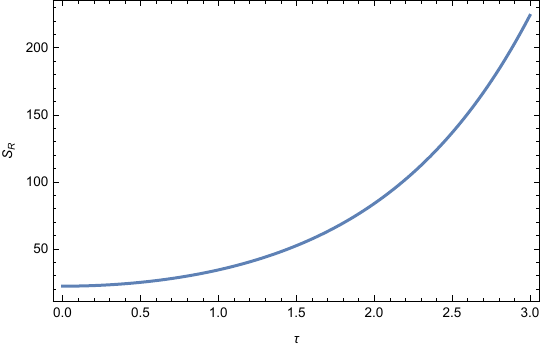}
    \caption{Time evolution of the radiation entropy in the no-island phase for $\rho_2 = 3$ and $\theta_0 = \pi/4$. The entropy grows monotonically with the boundary time $\tau$ and exhibits no sign of saturation.}
    \label{fig:noisland_entropy}
\end{figure}

We first consider the disconnected, no-island configuration. The bulk geometry $W_{d+1}$ corresponds to a portion of pure $\mathrm{AdS}_{d+1}$, with the line element
\begin{equation} \label{eq:ads_global_metric_no-island}
    ds^{2}_{d+1} = -\cosh^2 \rho_g\, d\tau_g^2 + d\rho_g^2 + \sinh^2 \rho_g\, d\Omega_{d-1}^2,
\end{equation}
where $\tau_{g}$ is the global AdS time, $r$ is the radial coordinate, and $d\Omega_{d-1}$ denotes the metric of $S^{d-1}$. The extremal hypersurface lies on a constant global time slice. The induced metric on such a slice as obtained from equation \eqref{eq:ads_global_metric_no-island} is given by
\begin{equation} \label{constant tau slice global metric}
    ds^2 = d\rho_g^2+\sinh^2 \rho_g \left(d\theta^2+\cos^2\theta\, d\Omega_{d-2}^2 \right).
\end{equation}
The corresponding area functional is
\begin{equation} \label{Area fb noisland phase}
    A = \Omega_{d-2} \int_{-\theta_0}^{\theta_0} d\theta\, \left(\sinh \rho_g\cos\theta \right)^{d-2} \sqrt{\rho_g'^2+\sinh^2 \rho_g},
\end{equation}
where $\Omega_{d-2}$ denotes the volume of the unit sphere $S^{d-2}$, and the prime denotes differentiation with respect to $\theta$.The extremal HRT surface in global AdS geometry is universal and independent of the bulk spacetime dimension. Its derivation is given in Appendix~\ref{app:RTsurface} (see Equation \eqref{RT_master}), and the resulting profile is
\begin{equation}
    \tanh \rho_g \cos\theta = \cos\theta_0 ,
\end{equation}
where $\theta_0$ specifies the boundary of the entangling region: $\theta \in \left[-\theta_{0}, \theta_{0} \right]$.Substituting this profile into Equation \eqref{Area fb noisland phase} eliminates the derivative $\rho_g'(\theta)$ and reduces the area functional to the single integral \footnote{To regulate the UV divergence at the asymptotic AdS boundary, we introduce an angular cutoff $\epsilon$, so that the RT surface terminates at $\theta=\pm(\theta_0-\epsilon)$ instead of $\theta=\pm\theta_0$.}
\begin{equation}
    A = \Omega_{d-2}\, \sin\theta_0 \cos^{d-1}\theta_0 \int_{-\theta_0+\epsilon}^{\theta_0-\epsilon} d\theta\, \frac{\cos^{d-2}\theta}{\left(\cos^2\theta-\cos^2\theta_0 \right)^{d/2}} .
\end{equation}
The detailed evaluation of this integral is presented in Appendix \ref{appendix:NoIslandArea}. Performing the integration yields
\begin{equation}\label{eq:area fn result no island phase}
    A = 2\Omega_{d-2} \cot^{d-1}\theta_0\,\sin \left(\theta_0-\epsilon \right)\, F_1 \left(\frac{1}{2}, -\frac{d-3}{2}, \frac{d}{2}, \frac{3}{2}, \sin^2 \left(\theta_0 - \epsilon \right), \sin^2 \left(\theta_0 - \epsilon \right) \csc^2\theta_0 \right),
\end{equation}
where $F_1$ denotes the Appell hypergeometric function of two variables \cite[\href{https://dlmf.nist.gov/16.13}{(16.13)}]{NIST:DLMF}. To obtain the time dependence of the entropy, we express the UV cutoff parameter $\epsilon$ in terms of the de Sitter global time $\tau$ appearing in the dS slicing of AdS. Throughout this subsection, $\tau_g$ denotes the global AdS time appearing in the metric \eqref{eq:ads_global_metric_no-island}, while $\tau$ denotes the boundary de Sitter global time related to the cutoff surface through the coordinate transformation \eqref{eq:coordinate_map}.Evaluating the RT surface profile, given in Equation \eqref{RT_master}, at the cutoff surface $\rho_g=\rho_t$ yields
\begin{equation}
    \cos\left(\theta_0-\epsilon(\tau)\right)
    =
    \frac{\cos\theta_0}{\tanh \rho_t(\tau)} .
\end{equation}
The cutoff radius $\rho_t$ is related to the dS-slicing coordinates through the coordinate transformation given in Equation \eqref{eq:coordinate_map},
\begin{equation}
    \rho_t(\tau)
    =
    \sinh^{-1}\!\left(\sinh\rho_c\,\cosh\tau\right).
\end{equation}
Therefore we obtain the time-dependent cutoff parameter as
\begin{equation}\label{eq:coutoff relation wit ds time}
    \epsilon(\tau)
    =
    \theta_0
    -
    \cos^{-1}\!\left(
    \frac{\cos\theta_0}
    {\tanh\!\left[
    \sinh^{-1}\!\left(\sinh\rho_c\,\cosh\tau\right)
    \right]}
    \right).
\end{equation}
Substituting this relation \eqref{eq:coutoff relation wit ds time} into the area functional result \eqref{eq:area fn result no island phase} , we obtain the time-dependent no-island entropy
\begin{equation}
    S_R(\tau) = \frac{A(\epsilon(\tau))}{4G_N}.
\end{equation}
Using the cutoff relation \eqref{eq:coutoff relation wit ds time} derived above, we numerically evaluate the no-island entropy as a function of the boundary time $\tau$. The resulting behavior is shown in Figure \ref{fig:noisland_entropy}. The entropy increases monotonically with time, reflecting the continuous growth of entanglement between the radiation and its complement in the absence of an island contribution.
The radiation entropy becomes large at late times. Since no island contribution is included in this phase, the entropy does not exhibit a Page transition or late-time saturation. This behavior motivates the inclusion of island configurations, which we discuss in the next.

\subsection{Island Phase}

\begin{figure}[t]
    \centering
    \begin{subfigure}[t]{0.50\textwidth}
        \centering
        \includegraphics[width=\textwidth]{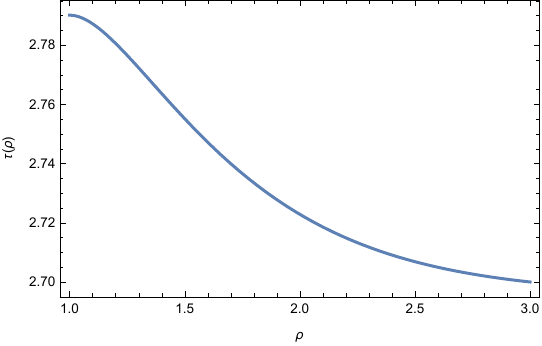}
        \caption{}
        \label{fig:Island_RT_reg_soln}
    \end{subfigure}
    \hfill
    \begin{subfigure}[t]{0.4\textwidth}
        \centering
        \includegraphics[width=\textwidth]{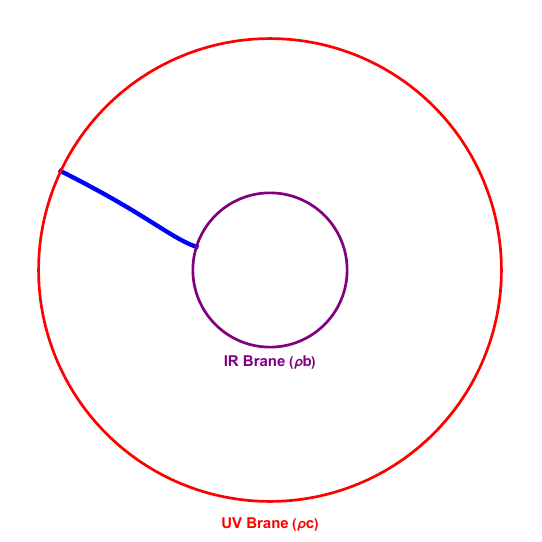}
        \caption{}
        \label{fig:Island_RT_polar_soln}
    \end{subfigure}
    \caption{Numerical RT surface in the island phase. Left: extremal surface profile $\tau(\rho)$ . Right: polar visualization of the same RT surface in the dS wedge bulk geometry.}
    \label{fig:Island RT both plot}
\end{figure}
We now turn to the connected extremal surfaces, which correspond to the island phase of the radiation entropy. In this configuration, the RT surface extends from the UV brane $Q_{2}$ and terminates on the IR brane $Q_1$. The existence of such surfaces is crucial for recovering a unitary Page curve at late times (see e.g. \cite{Almheiri:2019hni,Penington:2019npb,Almheiri:2019psf,Hartman:2020swn,Hashimoto:2020cas} and references therein). To study these surfaces, we employ the AdS coordinates adapted to dS slicing \eqref{Ads_global_ds_slice_metric}. Owing to the symmetry of the boundary subregion $\theta\in[-\theta_0,\theta_0]$, the extremal surface is completely characterized by its profile in the $(\rho,\tau)$ plane, allowing us to restrict the analysis to the constant angular slice $\theta=\theta_0$. The bulk geometry then takes the form
\begin{equation}
    ds^2 = d\rho^2 + \sinh^2\rho \left(- d\tau^2 + \cosh^2\tau \cos^2\theta_0\, d\Omega_{d-2}^2 \right).
\end{equation}
Describing the extremal surface by the embedding function $\tau=\tau(\rho)$, the induced geometry yields the area functional
\begin{align}
    A = 2\, \Omega_{d-2} \left(\cos\theta_0)^{d-2} \int d\rho \left(\sinh\rho\cosh\tau \left(\rho \right) \right)^{d-2} \sqrt{1-\sinh^2\rho\, \tau' \left(\rho \right)^2} \right).
\label{Area fn for HEE in connected phase}
\end{align}
The prefactor of 2 reflects the symmetry of the setup and accounts for the pair of identical extremal surfaces attached to the two endpoints of the radiation region as shown in Figure \ref{RT on island phase}. The extremal profile is determined by varying the functional \eqref{Area fn for HEE in connected phase}. This leads to the nonlinear differential equation
\begin{equation}
    \begin{split}
        &\mathrm{csch}(\rho) \left(d\,\coth(\rho)\,\tau' + \tau'' \right) + (d-2)\,\mathrm{csch}^{3}(\rho)\,\tanh(\tau)\\
        &\hspace{1.4 cm} - (d-2)\,\mathrm{csch}(\rho)\, \tanh(\tau)\, \tau'^2 - (d-1)\, \cosh(\rho)\, \tau'^3 = 0.
    \end{split}
\end{equation} \label{IslandEOM}
where primes denote derivatives with respect to $\rho$.

For dimensions $d>2$, we have not been able to obtain an analytic solution to equation \eqref{IslandEOM}. Consequently, the extremal surface must be determined numerically. In the present analysis, the surface is anchored on the UV brane while satisfying an orthogonality condition at the IR brane,
\begin{equation} \label{IslandBC}
    \tau \left(\rho_2 \right) = \tau_{\mathrm{UV}},\quad \tau' \left(\rho_1 \right) = 0.
\end{equation}
As an illustrative example, we choose
\begin{equation}
    \rho_1 = 1,\quad \rho_2 = 3,\quad \tau_{\mathrm{UV}} = 2.7,\quad d = 3,\quad \theta_0=\frac{\pi}{4}.
\end{equation}
The corresponding numerical solution is displayed in Figure \ref{fig:Island_RT_reg_soln}. For a clearer geometric interpretation, the same solution is shown in polar coordinates in Figure \ref{fig:Island_RT_polar_soln}. One observes that the surface leaves the UV brane smoothly and reaches the IR brane with vanishing slope, in agreement with the Neumann boundary condition.

Evaluating the area functional \eqref{Area fn for HEE in connected phase} on the numerical solution gives
\begin{equation}
    S_R^{(\mathrm{Island})} = \frac{A}{4G_{N}} \simeq 142.795 = \mathrm{Constant},
\end{equation}
where we have set $G_{N}=1$.

\subsection{Page curve}

A notable property of the connected phase is that its entropy remains finite and does not exhibit secular growth with the boundary time. This behavior is markedly from the disconnected saddle, whose entropy continues to increase as the system evolves. As a result, the connected extremal surface eventually becomes the preferred saddle and governs the late-time behavior of the radiation entropy. This transition is responsible for the emergence of the Page curve and the restoration of unitarity in the holographic description.
\begin{figure}[t]
    \centering 
    \includegraphics[width =0.7\textwidth]{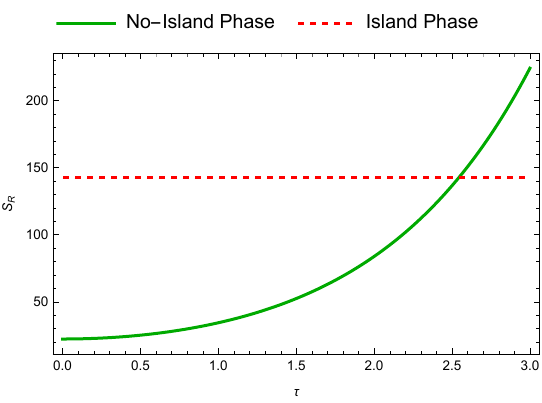} 
    \caption{Page curve obtained from the competition between the no-island and island phases. The entropy initially follows the no-island solution and grows with time. After the Page time, the island contribution dominates, causing the entropy to saturate.}
    \label{pagecurve}
\end{figure}
To investigate the role of the island contribution in restoring unitarity, we compare the entanglement entropy in the no-island and island phases. The resulting Page curve is shown in Figure \ref{pagecurve}. This behavior is consistent with the expected unitary evolution of the system.

\section{Discussions and concluding remarks} \label{sec:conclusion}

In this work, we studied the effective gravitational dynamics and quantum-information quantities in de Sitter wedge holography. Our setup consists of a region of $\mathrm{AdS}_{d+1}$ bounded by two end-of-the-world branes with induced $\mathrm{dS}_{d}$ geometries. The
codimension-two ends of the wedge are spheres, on which the proposed $\mathrm{CFT}_{d-1}$ dual is defined.

We first studied the effective description on the branes. After integrating over the radial direction of the wedge, we found an Einstein gravity action on the de Sitter slices. The effective Newton constant depends on the positions of the two branes through an integral. This
gives a direct relation between the higher-dimensional wedge description and the lower-dimensional gravitational description on the branes.

We then evaluated the gravitational partition function for $d=3,4,$ and $5$. In even dimensions of the defect CFT, the logarithmic divergence allowed us to extract the corresponding central charge. The central charges
are imaginary. This is consistent with the expectation that the CFT associated with de Sitter holography is generally non-unitary \cite{Strominger:2001pn, Maldacena:2002vr}. We obtained the same central charges from the universal part of the holographic entanglement entropy. The agreement between these two independent computations provides a useful check of the proposed holographic dictionary.

The spectrum of gravitational fluctuations gives further information about the effective theory. For Dirichlet boundary conditions, we did not find a massless graviton mode. In contrast, Neumann boundary conditions admit a
massless mode together with a tower of massive modes. The Schr\"{o}dinger form of the fluctuation equation gives a volcano-type potential, and the zero-mode wavefunction is concentrated near the outer brane. These results
show that the choice of boundary conditions is important for the appearance of localized gravity in the de Sitter wedge.

We also checked the first law of entanglement entropy. Orthogonality of the radial wavefunctions implies that the massive graviton modes do not change the holographic entanglement entropy at first order. The same modes
also do not contribute to the expectation value of the modular Hamiltonian. The first
law is therefore satisfied trivially for massive fluctuations. For the massless mode, we evaluated the two variations explicitly and found that they agree. The negative variation of the modular Hamiltonian is consistent with the unusual properties of the proposed boundary theory and, together with the imaginary central charges, provides another indication of its non-unitary character.

The non-unitary character of the proposed boundary CFT also raises an interesting question concerning the relation between the holographic entanglement entropy studied here and pseudo entropy \cite{Nakata:2020luh, Mollabashi:2020yie, Mollabashi:2021xsd}. In the present work,
we have used the standard Ryu-Takayanagi prescription, which is available because the ambient $(d+1)$-dimensional geometry is asymptotically AdS. Nevertheless, since the codimension-two boundary theory is expected to be non-unitary, it is not clear whether its field-theoretic interpretation should always be that of an ordinary
entanglement entropy or a pseudo entropy based on a transition matrix. We have not considered any such connection in the present work, and leave this question for future study.

Another quantity considered in this work is the timelike entanglement entropy between the two codimension-two defects. We constructed a connected extremal surface joining timelike separated regions of the defects. The surface has a Lorentzian induced metric everywhere and therefore gives a
purely imaginary area. In the example studied numerically, this leads to a purely imaginary timelike entropy. For nearby branes, the connected surface is expected to dominate because the two defects remain strongly correlated.
For larger brane separations, it will be important to compare its area with that of the disconnected surfaces. Such a comparison may reveal a new phase transition in the timelike entropy.

Finally, we studied a Page-like transition. We treated the infrared brane as a gravitating sector and moved the ultraviolet brane close to the AdS boundary so that it behaves as a non-gravitating bath. The disconnected
extremal surface gives a radiation entropy that increases with de Sitter time. The connected surface ending on the infrared brane gives an island contribution that remains finite. At sufficiently late times, the connected
surface has the smaller area, and the entropy changes from the growing no-island branch to the constant island branch. The resulting curve has the expected Page-curve form. Our analysis formulates this competition for general $d>2$, generalizing the work of \cite{Aguilar-Gutierrez:2023tic}, and illustrates it through numerical solutions.

There are some limitations to the present analysis. The background does not contain a dynamically evaporating black hole. The infrared brane only plays the role of an effective gravitating sector, while the expansion of the de
Sitter geometry changes the physical size of the radiation region. Thus, the Page curve found here should be viewed as an extremal-surface transition in the de Sitter wedge, rather than a complete model of black-hole evaporation. Moreover, the imaginary central charges indicate that the
defect theory is non-unitary. The saturation of the extremal-surface entropy therefore does not, by itself, establish the existence of a unitary microscopic de Sitter theory.

A natural extension of the present work is to introduce black holes in the bulk wedge geometry and on the de Sitter branes. Such a setup may provide a more concrete gravitational system in which the competition between the
island and no-island extremal surfaces can be studied. In particular, it may lead to a more direct mechanism for realizing the Page curve and for understanding information recovery in de Sitter wedge holography.

\section*{Acknowledgments}

This work is partially supported by the National Natural Science Foundation of China (NSFC) under grant no. 12247103. SP would like to thank Harvendra Singh and Ayan K. Patra for useful discussions. SP acknowledges the Saha Institute of Nuclear Physics (SINP) and the Department of Atomic Energy (DAE), Government of India, for financial support through a doctoral fellowship. The authors used ChatGPT (OpenAI) to assist with drafting and language editing of parts of the manuscript.

\appendix

\section{Some useful formulae} \label{geometric quantity}

In this appendix, we collect some intermediate results useful for the derivation in the main text. We start by rewriting the line element of $\left(d+1 \right)$-dimensional anti-de Sitter geometry in global coordinate system
\begin{equation} \label{eq:ads_global_metric_appendix}
    ds^{2}_{d+1} = -\cosh^2 \rho_g\, d\tau_g^2 + d\rho_g^2 + \sinh^2 \rho_g\, d\Omega_{d-1}^2,
\end{equation}
where $\tau_g$ is the global AdS time, $\rho_g$ is the radial coordinate, and $d\Omega_{d-1}$ denotes the metric of $(d-1)$-sphere with unit radius.

For the computations in this paper, we mainly use an AdS$_{d+1}$ metric foliated with dS$_{d}$ slices; which can be expressed as
\begin{equation} \label{eq:ads_dS_sliced_metric_appendix}
	ds^{2}_{d+1} = d\rho^{2} + \sinh^{2} \rho\, h_{ij}(x)dx^{i}dx^{j} \, ,
\end{equation}
where $h_{ij}(x)$, $\left(i, j = 1, 2, \ldots, d \right)$ denotes the $d$-dimensional de Sitter line element in any convenient coordinate system, with unit curvature radius. E.g. if we use global coordinates for the de Sitter slices, the complete metric is expressed as
\begin{equation} \label{eq:ads_ds_global_sliced_metric_appendix}
    ds_{d+1}^2 = d\rho^2 + \sinh^2\rho \left(-d\tau^2 + \cosh^2\tau\, d\Omega_{d-1}^2 \right).
\end{equation}
The coordinate $\tau \in (-\infty,\, \infty)$ is the global de Sitter time on each slice, and $\rho$ is the bulk radial coordinate of AdS. For any fixed value of $\rho$, the hypersurface $\rho = \mathrm{constant}$ is a
$d$-dimensional de Sitter space with curvature radius $\sinh\rho$. For convenience, we may express the metric on $S^{d-1}$ by foliating $S^{d-1}$ with an $S^{d-2}$ submanifold as
\begin{equation}
    d\Omega^{2}_{d-1} = d\theta^{2} + \cos^{2}\theta\, d\Omega^{2}_{d-2}.
\end{equation}
Here $\theta \in [-\frac{\pi}{2},\, \frac{\pi}{2}]$ denotes the angular coordinate on the sphere.

The coordinate systems in \eqref{eq:ads_global_metric_appendix} and \eqref{eq:ads_ds_global_sliced_metric_appendix} are related to each other by
\begin{equation} \label{eq:coordinate_map}
    \tan \tau_g = \sinh\tau \tanh \rho,\quad \sinh \rho_g=\sinh\rho \cos\tau.
\end{equation}

\subsection*{Background Tensors}

Starting from the metric \eqref{eq:ads_dS_sliced_metric_appendix}, we derive the corresponding background tensors.\\

\noindent \textit{Christoffel symbols:}
\begin{equation}
    \begin{split}
        \Gamma_{ij}^{\rho} = - \sinh(\rho) \cosh(\rho) h_{ij},\quad
        \Gamma_{\rho j}^{i} = \coth(\rho) \delta^{i}_{j},\quad
        \Gamma_{jk}^{i} = \Gamma_{jk}^{i\,\mathrm{(dS)}}.
    \end{split}
\end{equation}

\noindent \textit{Ricci tensor components:}
\begin{equation}
	R_{\rho \rho} = -d,\quad R_{ij} = R^{\mathrm{(dS)}}_{ij} -\left(\sinh^2(\rho)+(d-1)\cosh^2(\rho) \right) h_{ij}.
\end{equation}

\noindent \textit{Ricci scalar :}
\begin{equation} \label{Ricci scalar}
	\hat{R} = R^{\mathrm{(dS)}}\mathrm{csch^2(\rho)} - d\left(2 + \left(d - 1 \right) \coth^2 \left(\rho \right) \right).
\end{equation}

\subsection*{Extrinsic curvature of the brane}

Since the metric \eqref{eq:ads_dS_sliced_metric_appendix} is block-diagonal, the unit normal vector 
directed outward to each brane is given by
\begin{equation}
	\partial_{n} = \pm\partial_{\rho}.
\end{equation}
Therefore, the extrinsic curvature of each of the brane is
\begin{equation}
	K_{ij}=\frac{1}{2} \partial_{n} h_{ij} = \pm\coth(\rho_{a})\hat{h}_{ij},
\end{equation}
and its trace reads 
\begin{equation}
	\mathcal{K}=\pm d \coth(\rho_{a}).
\end{equation}
 The relative $\pm$ sign reflects that $n_{\mu}$ points in opposite directions on the two branes. In this convention, the positive sign corresponds to $Q_{2}$, while the negative sign corresponds to $Q_{1}$.
 
\subsection*{Bulk Einstein's field equations}

\subsubsection*{d = 2}
 
 From equation \eqref{eq:ads_ds_global_sliced_metric_appendix}, we can write the three-dimensional global AdS$_{3}$ spacetime with  dS$_{2}$ foliations as
 \begin{equation}
     dS_{3}^2 = d\rho^2+\sinh^2\rho\left(-d\tau^2+\cosh^2\tau \, d\theta^2\right)
 \end{equation}
 For this case, the non-zero Ricci tensor components are 
 \begin{equation} \label{eq:Ricci_tensor_3D_bulk}
     R_{\rho\rho}=-2,\quad R_{\tau\tau}=2\sinh^2\rho,\quad R_{\theta\theta}=-2\cosh^2\tau\sinh^2\rho,
 \end{equation}
 and the Ricci scalar is $R = -6$.
 
 The cosmological constant for three-dimensional Anti-de-Sitter spacetime is $\Lambda=-1$. Hence the bulk equation of motion is given by
 \begin{equation} \label{eq:bulk_eom_3D}
     R_{\mu\nu} - \frac{1}{2}g_{\mu\nu}R - g_{\mu\nu}=0,
 \end{equation}
 where $\mu,\nu\in\{\rho,\tau,\theta\}.$ One can check that bulk e.o.m. \eqref{eq:bulk_eom_3D} is satisfied for the Ricci tensors given in equation \eqref{eq:Ricci_tensor_3D_bulk}, and the Ricci scalar, $R = -6$.
 
\subsubsection*{d = 3}
 
Four-dimensional global AdS$_{4}$ foliated by dS$_{3}$ slices can be described with the following line element
\begin{equation}
     dS_{4}^2 = d\rho^2 + \sinh^2\rho \left(-d\tau^2 + \cosh^2\tau \left(d\theta^2 + \cos^2\theta \  d\Omega_{1}^2 \right) \right)
 \end{equation}
 The non-zero components of the Ricci tensor are 
 \begin{equation} \label{eq:Ricci_tensor_4D_bulk}
    \begin{split}
        R_{\rho\rho} &= -3,\\ R_{\tau\tau} &= 3\sinh^2\rho,\\ 
        R_{\theta\theta} &= -3 \cosh^2\tau \sinh^2\rho,\\
        R_{\phi\phi} &= -3 \cos^2\theta \cosh^2\tau \sinh^2\rho .
    \end{split}
 \end{equation}
 The  Ricci scalar and the cosmological constant, $\Lambda = -\dfrac{d\left(d-1\right)}{2}$, are easily obtained to be 
 \begin{equation}
     R = - 12,\quad \Lambda = -3.
 \end{equation}
 Hence the bulk e.o.m. is given as
 \begin{equation}\label{eq:bulk_eom_4D}
     R_{\mu\nu} - \frac{1}{2}g_{\mu\nu}R - 3 g_{\mu\nu}=0,
 \end{equation}
 where $\mu,\nu\in\{\rho,\tau,\theta,\phi\}.$ One can verify that the bulk equations of motion \eqref{eq:bulk_eom_4D} is indeed satisfied by the Ricci tensors given in \eqref{eq:Ricci_tensor_4D_bulk}, together with the Ricci scalar derived above.
 
In a similar manner, one can verify the bulk Einstein's field equations for other dimensions.
\section{Analytical structure of graviton mass spectrum} \label{appendix:analytical_graviton_ms}

In this appendix, we corroborate our findings regarding the graviton mass spectrum on the branes in Section \ref{sec2} with analytical arguments.

From Equation \eqref{eq:final mass spectrum equation}, the general solution of the radial equation is
\begin{equation}
H(\rho)
=
\operatorname{csch}\rho
\left(
c_{1}P_{1}^{\lambda}(\cosh\rho)
+
c_{2}Q_{1}^{\lambda}(\cosh\rho)
\right),
\qquad
\lambda=\sqrt{1-m^{2}}.
\label{eq:general_solution_appendix}
\end{equation}
Imposing the Dirichlet boundary condition \eqref{eq:BBC con} at
$\rho=\rho_i$ gives
\begin{equation}
H(\rho_i)=0
\quad\Longrightarrow\quad
c_{1}P_{1}^{\lambda}(x_i)
+
c_{2}Q_{1}^{\lambda}(x_i)
=
0,
\end{equation}
where $x_i=\cosh\rho_i,
\qquad
i=1,2.$ and $\tilde{c}=\frac{c_1}{c_2}$. 
Now, Using the associated Legendre function identities ( \cite[\href{https://dlmf.nist.gov/16.13}{§16.15}]{NIST:DLMF}),
\begin{align}
P_{\nu}^{-\mu}(x)
&=
(-1)^{\mu}
\frac{\Gamma(\nu-\mu+1)}
{\Gamma(\nu+\mu+1)}
P_{\nu}^{\mu}(x),
\\
Q_{\nu}^{-\mu}(x)
&=
(-1)^{\mu}
\frac{\Gamma(\nu-\mu+1)}
{\Gamma(\nu+\mu+1)}
Q_{\nu}^{\mu}(x),
\end{align}
and setting $\nu=1,
\mu=\lambda,$
we obtain
\begin{align}
P_{1}^{-\lambda}(x)
&=
(-1)^{\lambda}
\frac{\Gamma(2-\lambda)}
{\Gamma(2+\lambda)}
P_{1}^{\lambda}(x),
\\
Q_{1}^{-\lambda}(x)
&=
(-1)^{\lambda}
\frac{\Gamma(2-\lambda)}
{\Gamma(2+\lambda)}
Q_{1}^{\lambda}(x).
\end{align}

Therefore,
\begin{align}
\tilde{c}(-\lambda)=-\frac{Q_{1}^{-\lambda}(x_i)}
{P_{1}^{-\lambda}(x_i)}=
-
\frac{
(-1)^{\lambda}
\dfrac{\Gamma(2-\lambda)}
{\Gamma(2+\lambda)}
Q_{1}^{\lambda}(x_i)}
{
(-1)^{\lambda}
\dfrac{\Gamma(2-\lambda)}
{\Gamma(2+\lambda)}
P_{1}^{\lambda}(x_i)}=
-
\frac{Q_{1}^{\lambda}(x_i)}
{P_{1}^{\lambda}(x_i)}=
\tilde{c}(\lambda).
\end{align}
Therefore, the coefficient ratio is invariant under the transformation
$\lambda\rightarrow-\lambda$. Consequently, if $\lambda$ satisfies the Dirichlet boundary condition, then $-\lambda$ also satisfies the same boundary condition. An analogous analysis can be carried out for the Neumann boundary condition \eqref{eq:NBC con}. Owing to the different degrees of the associated Legendre functions appearing in the Neumann case, the derivation is algebraically more involved; however, our numerical analysis confirms the same symmetry under the transformation $\lambda\rightarrow-\lambda$.

To determine whether a massless graviton mode exists, we set $m=0$. The radial equation \eqref{eq:final mass spectrum equation} reduces to
\begin{equation}
\sinh^2\rho\,H''(\rho)
+
3\sinh\rho\cosh\rho\,H'(\rho)
=0.
\label{eq:massless_radial}
\end{equation}
Writing Equation \eqref{eq:massless_radial} as
\begin{equation}\label{eq:diff form}
\frac{d}{d\rho}
\left(
\sinh^3\rho\,H'
\right)
=0,
\end{equation}
its general solution is
\begin{equation}
H(\rho)
=
A
+
B
\left[
\log\!\left(\tanh\frac{\rho}{2}\right)
-
\operatorname{csch}\rho\,\operatorname{coth}\rho
\right],
\label{eq:massless_solution}
\end{equation}
where $A$ and $B$ are integration constants.Imposing the Dirichlet boundary conditions \eqref{eq:BBC con},
\[
H(\rho_1)=H(\rho_2)=0,
\]
gives
\begin{align}
A+B\,F(\rho_1)=0 \ \mathrm{and} \ A+B\ F(\rho_2)=0,
\end{align}
where
\begin{equation}
F(\rho)
=
\log\!\left(\tanh\frac{\rho}{2}\right)
-
\operatorname{csch}\rho\,\operatorname{coth}\rho.
\end{equation}

Since $F(\rho)$ is not constant on the interval
$\rho\in[\rho_1,\rho_2]$, one has $F(\rho_1)\neq F(\rho_2),$
which immediately implies $A=B=0.$ Hence,
\[
H(\rho)\equiv0,
\]
is the only solution satisfying the Dirichlet boundary conditions. Therefore, $m=0$ is not an eigenvalue of the Dirichlet problem, and consequently the Dirichlet spectrum does not contain a massless graviton mode.

Again, for the Neumann boundary condition \eqref{eq:NBC con}, from Equation \eqref{eq:diff form} the general solution satisfies
\begin{equation}
H'(\rho)=\frac{C}{\sinh^3\rho},
\label{eq:Hprime_massless}
\end{equation}
where $C$ is an integration constant. Imposing the Neumann boundary conditions \eqref{eq:NBC con} ,$H'(\rho_1)=H'(\rho_2)=0$, gives
\begin{align}
\frac{C}{\sinh^3\rho_1}=0,
\quad
\text{and}
\quad
\frac{C}{\sinh^3\rho_2}=0.
\end{align}
Since $\sinh\rho>0$ for $\rho\in[\rho_1,\rho_2]$, it immediately follows that
$C=0$.Hence, $H'(\rho)=0$,which implies
\[
H(\rho)=A,
\]
where $A$ is an arbitrary constant. Therefore, for $A\neq0$, the solution is non-trivial and automatically satisfies the Neumann boundary conditions \eqref{eq:NBC con}. Consequently, $m=0$ is an eigenvalue of the Neumann problem, and hence the Neumann spectrum contains a genuine massless graviton mode.
\section{Vector field on the brane} \label{appendix:vector_spectrum}

In this section, we investigate the mass spectrum of vector fields on the end-of-the-world branes. The analysis is very similar to that of the graviton sector. For simplicity, we focus on the case $d+1=4$. In the probe limit, the bulk geometry and the locations of the branes are given by
\begin{subequations}
    \begin{align}
        ds^2 &= d\rho^2+\sinh^2(\rho)h^{(0)}_{ij}(x)dx^{i}dx^{j}, \label{perturbed_metric} \\
        Q_1 &: \rho=\rho_1,
        \quad
        Q_2 : \rho=\rho_2 .
    \end{align}
\end{subequations}
Here $h^{(0)}_{ij}$ denotes the de Sitter metric with unit radius. We consider the following ansatz for the bulk Maxwell field
\begin{equation}
\mathcal{A}_{\rho} = 0,
\quad
\mathcal{A}_{i} = S(\rho)A_{i}(x).
\label{vectorans}
\end{equation}
Substituting into the Maxwell equation
\begin{equation}
\nabla_{\mu}\mathcal{F}^{\mu\nu}=0,
\end{equation}
and separating the $\rho$ and $x$ dependence, we obtain
\begin{equation}
\bar{D}_{i}F^{ij}-m_v^2A^{j}=0,
\end{equation}
together with the radial equation
\begin{equation}\label{rad eq 2}
\sinh^2(\rho)S''(\rho)
+\sinh(\rho)\cosh(\rho)\,S'(\rho)
+m_v^2S(\rho)=0.
\end{equation}
Here
\begin{equation}
F_{ij}=\bar{D}_{i}A_{j}-\bar{D}_{j}A_{i},
\end{equation}
where $\bar{D}_{i}$ is the covariant derivative associated with the background metric $h^{(0)}_{ij}$, and $m_v$ denotes the mass of the vector mode on the brane.

For $m_{v} \neq 0$, the general solution to \eqref{rad eq 2} is found to be given by
\begin{equation} \label{rad soln of radeq 2}
    S(\rho) = C_{1} \cos\left(m_v\log\left(\tanh \frac{\rho}{2}\right) \right) + C_{2} \sin\left(m_v\log\left(\tanh \frac{\rho}{2} \right) \right).
\end{equation}

The massless modes should be treated separately. For $m = 0$, we find that the general solution to the radial equation is
\begin{equation} \label{eq:rad_soln_of_radeq_2_massless}
    S\left(\rho \right) = C_{1} + C_{2} \log \left(\tanh \frac{\rho}{2} \right).
\end{equation}
As usual, $C_{1}$ and $C_{2}$ are two constants of integration -- which can be determined by analysing the boundary conditions. 

\subsection*{Dirichlet Boundary condition}

The Dirichlet boundary condition for a vector field in the wedge geometry reads
\begin{equation}
    S(\rho_1)=0,\quad S(\rho_2) = 0
\end{equation}
For the massive modes given by equation \eqref{rad soln of radeq 2}, these conditions are equivalent to the vanishing of the ``spectral function"
\begin{align} \label{eq:Dirichlet_bc_vector}
    D_{v}^{\mathrm{dS}}(m_v,\rho_1,\rho_2) = \sin\left(m_v \left(\log\tanh\frac{\rho_2}{2} - \log\tanh\frac{\rho_1}{2} \right) \right) = 0.
\end{align}
For notational convenience, let us define
\begin{equation} \label{eq:Delta_definition}
    \Delta = \log \left(\tanh \frac{\rho_2}{2} \right) - \log \left(\tanh \frac{\rho_1}{2} \right),\quad \rho_2 > \rho_1 .
\end{equation}
Since $\tanh(\rho_2/2) > \tanh(\rho_1/2)$, one has $\Delta > 0$.\\

\noindent Therefore, we find that condition \eqref{eq:Dirichlet_bc_vector} is satisfied for  
\begin{equation}
    m_v^{(n)} = \frac{n\pi}{\Delta},\quad n\in\mathbb Z.
\label{DirichletRoots}
\end{equation}
The roots occur symmetrically around the origin,
\begin{equation*}
\cdots,
-\frac{3\pi}{\Delta},
-\frac{2\pi}{\Delta},
-\frac{\pi}{\Delta},
0,
\frac{\pi}{\Delta},
\frac{2\pi}{\Delta},
\frac{3\pi}{\Delta},
\cdots .
\end{equation*}
This symmetry follows from
\begin{equation*}
D_{v}^{\mathrm{dS}} \left(-m_v \right) = -D_{v}^{\mathrm{dS}} \left(m_{v} \right).
\end{equation*}
However, since both the radial equation and the effective vector equation on the brane depend only on \(m_v^2\), the roots \(m_v\) and \(-m_v\) correspond to the same physical KK state. Consequently, the physical spectrum is conventionally restricted to the non-negative branch.

From equation \eqref{DirichletRoots}, the spacing between consecutive Kaluza-Klein (KK) levels is
\begin{equation}
\delta m_v = m_v^{(n+1)} - m_v^{(n)} = \frac{\pi}{\Delta},
\end{equation}
which is independent of \(n\). Therefore, the Dirichlet spectrum forms an exactly equally spaced KK tower.

For the massless modes in equation \eqref{eq:rad_soln_of_radeq_2_massless}, imposing the Dirichlet boundary conditions on both branes forces
\begin{equation}
C_1 = C_2 = 0.
\end{equation}
Hence, only the trivial solution exists, and no massless vector mode is present in the Dirichlet spectrum.
\begin{figure}[t]
    \centering
    \begin{subfigure}{0.48\textwidth}
        \centering
        \includegraphics[width=\textwidth, height=5.5 cm]{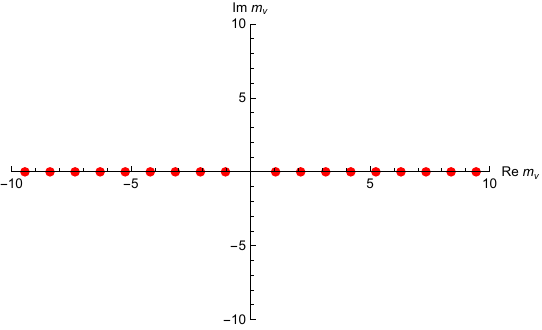}
        \label{fig:dirichlet_zeros_vector}
        \caption{}
    \end{subfigure}
    \hfill
    \begin{subfigure}{0.48\textwidth}
        \centering
        \includegraphics[width=\textwidth]{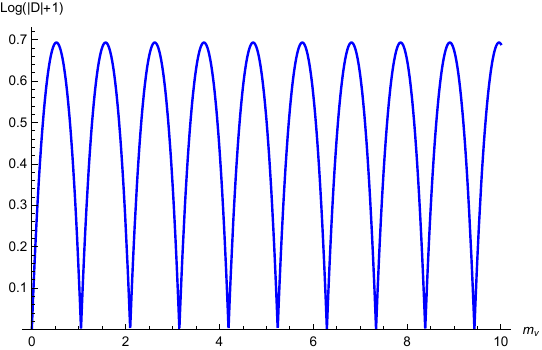}
        \caption{}
        \label{fig:logD_vector_2}
    \end{subfigure}
    \caption{(a) Zeros of the Dirichlet spectral function $D_{v}^{\mathrm{dS}} \left(m_{v}, \rho_{1}, \rho_{2} \right)$, (b) Log plot of $D_{v}^{\mathrm{dS}} \left(m_{v}, \rho_{1}, \rho_{2} \right)$ for $\rho_{1} = 0.1$ and $\rho_{2} = 10$; the downward pointing spikes denote the zero points.}
    \label{fig:Dirichlet_plot_vector}
\end{figure}

\subsection*{Neumann Boundary condition}

Next, we consider the situation where we impose Neumann Boundary condition on the two ETW dS branes
\begin{equation}
    \partial_{\rho}S(\rho_1)=0,\quad \partial_{\rho}S(\rho_2)=0.
\end{equation}
After some straightforward algebra, we find that the Neumann boundary condition is equivalent to
\begin{align} \label{eq:Neumann_bc_vector}
N_{V}^{\mathrm{dS}} \left(m_v, \rho_1, \rho_2 \right) &=
m_v^2
\,\mathrm{csch}(\rho_1)\,\mathrm{csch}(\rho_2) \sin \left(m_v \Delta \right) = 0,
\end{align}
where $\Delta$ is defined according to equation \eqref{eq:Delta_definition}. For \(m_v\neq0\), equation \eqref{eq:Neumann_bc_vector} reduces to
\begin{equation}
    \sin \left(m_v\Delta \right) = 0,
\end{equation}
which leads to the solution
\begin{equation}
    m_v^{(n)} = 
\frac{n\pi}{\Delta},
\qquad
n=\pm1,\pm2,\ldots .
\end{equation}
Hence the massive Neumann spectrum is exactly identical to the Dirichlet spectrum and exhibits the same uniform spacing
\begin{equation*}
    \delta m_v = \frac{\pi}{\Delta}.
\end{equation*}
The overall factor $m_v^2$ in equation \eqref{eq:Neumann_bc_vector} implies that $m_v=0$ is automatically a root of the spectral equation. In this case, imposing the Neumann boundary conditions on the general solution \eqref{eq:rad_soln_of_radeq_2_massless} yields \[C_2=0, \] and hence
\begin{align}
    S(\rho) = C_1,
\end{align}
which represents a genuine non-trivial massless vector mode on the brane. The appearance of the zero mode distinguishes the Neumann spectrum from the Dirichlet one.
\begin{figure}[t]
    \centering
    \begin{subfigure}[t]{0.48\textwidth}
        \centering
        \includegraphics[width=\textwidth]{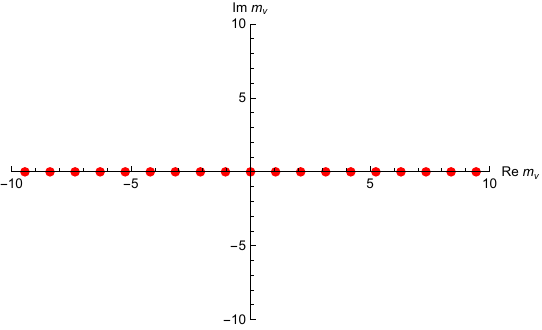}
        \caption{}
        \label{fig:neumann_zeros_vector}
    \end{subfigure}
    \hfill
    \begin{subfigure}[t]{0.48\textwidth}
        \centering
        \includegraphics[width=\textwidth]{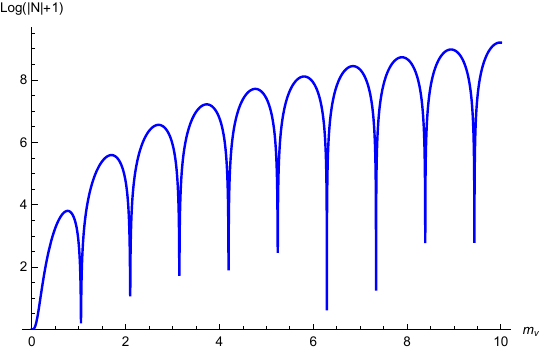}
        \caption{}
        \label{fig:log_neumann_zeros_vector}
    \end{subfigure}
    \caption{(a) Roots of the Neumann spectral function $\left|N_{v}^{\mathrm{dS}} \left(m_v, \rho_{1}, \rho_{2} \right) \right|$, (b) Log plot of $\left|N_{v}^{\mathrm{dS}} \left(m_{v}, \rho_{1}, \rho_{2} \right) \right|$ for $\rho_{1} = 0.1$ and $\rho_{2} = 10$; the downward pointing spikes denote the zero points.}
    \label{fig:Neumann_plot_vector}
\end{figure}

Please refer to the figures \ref{fig:Dirichlet_plot_vector} and \ref{fig:Neumann_plot_vector} for a graphical representation of the mass spectrum for Dirichlet and Neumann b.c., respectively.

\subsection*{Possibility of tachyonic instability}

To check for the possibility of imaginary mass modes, let
\begin{equation*}
    m_v=i\mu,\quad \mu\in\mathbb R.
\end{equation*}
Then the massive spectral conditions \eqref{eq:Dirichlet_bc_vector} and \eqref{eq:Neumann_bc_vector} become
\begin{subequations}
    \begin{align}
        \mathrm{Dirichlet:}\qquad D_{v}^{\mathrm{dS}} \left(i\mu, \rho_1, \rho_2 \right) &= i\sinh\mu\Delta.\\
        \mathrm{Neumann:}\qquad N_{v}^{\mathrm{dS}}\left(i\mu,\rho_1,\rho_2 \right) &= -i \mu^2\, \mathrm{csch}\left(\rho_1 \right) \,\mathrm{csch}\left(\rho_2 \right) \sinh \left(\mu\Delta \right).
    \end{align}
\end{subequations}

Since \(\Delta>0\), the spectral equation in both cases admits only the trivial solution \(\mu=0\). Therefore, no non-trivial imaginary roots exist, implying that all eigenvalues are real and satisfy $m_v^2 \geq 0$. The vector modes in the wedge are, therefore, free from tachyonic instability.
\subsection*{Large-separation limit}

In the large-separation limit, where
\begin{equation}
\rho_1\ll1,
\qquad
\rho_2\gg1,
\end{equation}
we may use the asymptotic approximations
\begin{equation}
\tanh\frac{\rho_1}{2}\simeq\frac{\rho_1}{2},
\qquad
\tanh\frac{\rho_2}{2}\simeq1,
\end{equation}
which imply
\begin{equation}
\Delta
\simeq
\log\frac{2}{\rho_1}.
\end{equation}
The KK masses therefore become
\begin{equation}
m_v^{(n)}
\simeq
\frac{n\pi}{\log(2/\rho_1)}.
\end{equation}
As the brane separation increases, $\Delta$ grows and the KK masses decrease. Consequently, the spacing
\begin{equation}
\delta m_v=\frac{\pi}{\Delta}
\end{equation}
becomes progressively smaller. In the limit
\begin{equation}
\rho_1\rightarrow0,
\qquad
\rho_2\rightarrow\infty,
\end{equation}
one finds
\begin{equation}
\delta m_v\rightarrow0,
\end{equation}
and the discrete KK spectrum approaches a continuum.

\section{Ryu-Takayanagi (RT) hypersurface in global AdS$_{d+1}$} \label{app:RTsurface}

In this appendix, we derive the RT surface associated with a spherical entangling region in $\left(d+1 \right)$-dimensional anti-de Sitter geometry in global coordinates. We show that the shape of the extremal surface is independent of the spacetime dimensions.

\subsection*{AdS$_{d+1}$ as a hyperboloid}

Unit-radius AdS$_{d+1}$ can be represented as the hyperboloid
\begin{equation} \label{AdS_hyperboloid}
-X_{-1}^{2}-X_{0}^{2} + \sum_{i=1}^{d}X_i^{2} = -1,
\end{equation}
embedded in the flat space $\mathbb{R}^{2,d}$ with metric
\begin{equation}
    ds_{\rm emb}^{2} = -dX_{-1}^{2} - dX_{0}^{2} + \sum_{i=1}^{d} dX_i^{2}.
\end{equation}
We consider a constant global-time slice, $t_g=0$, for which $X_{0} = 0$. The remaining embedding coordinates are parameterized by global coordinates $(\rho_g, \theta, \Omega_{d-2})$ as
\begin{align} \label{eq:ads_embeddin_in_R}
    \begin{split}
        X_{-1} &= \cosh\rho_g ,\\
        X_1 &= \sinh\rho_g \cos\theta,\\
        X_2 &= \sinh\rho_g \sin\theta \cos\phi_1,\\
        X_3 &= \sinh\rho_g \sin\theta \sin\phi_1 \cos\phi_2,\\
        &\hspace{0.5cm}\vdots \\
        X_d &= \sinh\rho_g \sin\theta \sin\phi_1 \ldots \sin\phi_{d-2}.
    \end{split}
\end{align}
Substituting these coordinate into equation \eqref{AdS_hyperboloid} immediately reproduces the hyperboloid constraint.
%

\subsection*{RT hypersurface in embedding space}

Let us consider a spherical cap on the asymptotic boundary bounded by
\begin{equation*}
    \theta = \pm \theta_0 .
\end{equation*}
The entangling region preserves an $SO(d-1)$ rotational symmetry acting on the transverse sphere $S^{d-2}$. The corresponding RT surface maintains the same symmetry. Therefore, the extremal surface can only depend on the coordinates $(\rho_g, \theta)$.

For a spherical entangling region in the vacuum state of a holographic CFT, the RT surface is a totally geodesic submanifold of a constant time slice of AdS \cite{Ryu:2006bv, Casini:2011kv}. In embedding space, totally geodesic submanifolds arise as intersections of the AdS hyperboloid with linear subspaces passing through the origin \cite{Ratcliffe:2006bfa}. Consider a general hyperplane
\begin{equation}
V_M X^M =0 .
\end{equation}
Since the RT surface preserves the full rotational symmetry of the transverse sphere, the normal vector $V_{M}$ must also be invariant under $SO \left(d-1 \right)$
\begin{equation*}
V_2=V_3=\cdots =V_d =0 .
\end{equation*}
The hyperplane equation therefore reduces to
\begin{equation}
X_{-1} = C X_1,\quad C \equiv -\frac{V_{1}}{V_{-1}}.
\end{equation}
Substituting the embedding coordinates \eqref{eq:ads_embeddin_in_R} into the last equation yields
\begin{equation} \label{intermediate_relation}
    \tanh\rho_g \cos\theta = \frac{1}{C}.
\end{equation}

\noindent Near the asymptotic AdS$_{d+1}$ boundary $\rho_g \rightarrow \infty$, we have $\tanh\rho_g \rightarrow 1$. At the asymptotic boundary, the extremal surface terminates at $\theta=\pm \theta_0$. Using this boundary condition in equation \eqref{intermediate_relation} we obtain
\begin{equation}
    1\times \cos\theta_0 = \frac{1}{C}.
\end{equation}
Therefore, the RT hypersurface for a spherical entangling region in global AdS$_{d+1}$ is shown to be given by
\begin{equation}
    \tanh\rho_g \cos\theta = \cos\theta_0.
\label{RT_master}
\end{equation}
The hypersurface can also be expressed in the following equivalent forms
\begin{align}
    \rho_g(\theta) &= \operatorname{arcsinh} \left(\frac{\cos\theta_0}{\sqrt{\cos^2\theta-\cos^2\theta_0}} \right).
    \rho_g(\theta) &= \operatorname{arcsinh} \left(\frac{\cos\theta_0}{\sqrt{\sin^2\theta_0 \cos^2\theta - \cos^2\theta_0 \sin^2\theta}} \right). \label{eq:RT_surface_solution_alt2}
\end{align}
As a consistency check, for $d=2$, direct extremization of the geodesic length functional exactly matches with equation \eqref{eq:RT_surface_solution_alt2}.

At no stage in the derivation does the explicit value of $d$ enter. The dimension of spacetime enters only through the rotational symmetry group acting on the transverse sphere $S^{d-2}$. Consequently the geometric profile of the RT surface is universal.

\subsection*{Relation to the Poincar\'e hemisphere}

It is instructive to verify that the RT surface obtained above is equivalent to the familiar hemisphere solution in the Poincar\'e patch of AdS$_{d+1}$. This provides an alternative derivation of the dimension independence of the extremal surface, and establishes direct contact with the standard Ryu-Takayanagi construction. Using the embedding coordinates in equation \eqref{eq:ads_embeddin_in_R}, the RT surface equation \eqref{RT_master} can be rewritten as
\begin{equation}
\tanh\rho_g \cos\theta
=
\frac{\sinh\rho_g\cos\theta}{\cosh\rho_g}
=
\frac{X_1}{X_{-1}}
=
\cos\theta_0.
\label{RT_embedding_form}
\end{equation}

We now introduce the Poincar\'e coordinates \((z,r,\Omega_{d-2})\) in the constant-time slice of AdS. In terms of the embedding coordinates, these are given by
\begin{subequations}
\begin{align}
X_{-1}
&=
\frac{1}{2z}
\left(
1+z^2+r^2
\right), \\
X_{1}
&=
\frac{1}{2z}
\left(
r^2+z^2-1
\right).
\end{align}
\end{subequations}
The remaining coordinates are proportional to the Cartesian coordinates on \(\mathbb{R}^{d-1}\). Substituting these expressions into equation\eqref{RT_embedding_form}, we obtain
\begin{equation}
\frac{r^2+z^2-1}
     {r^2+z^2+1}
=
\cos\theta_0 .
\end{equation}
Using the trigonometric identity
\begin{equation}
\frac{1+\cos\theta_0}
     {1-\cos\theta_0}
=
\cot^2\frac{\theta_0}{2},
\end{equation}
we observe that the RT surface equation takes the form
\begin{equation}
r^2+z^2
=
R^2,
\qquad
R
=
\cot\frac{\theta_0}{2}.
\label{hemisphere_RT}
\end{equation}

Equation \eqref{hemisphere_RT} is precisely the standard hemisphere solution describing the RT surface of a spherical entangling region in Poincar\'e AdS.

\section{Extremal surface and entanglement entropy in dS wedge holography} \label{appendix:RT_hypersurface_dS}

We provide here details of the computation of holographic entanglement entropy in wedge holography with de Sitter branes. Our exposition here is heavily inspired by \cite{Ogawa:2022fhy}. We consider the $d$-dimensional de Sitter geometry in two coordinate systems
\begin{subequations}
    \begin{align}
        \text{Global coordinates:}\quad ds^2 &= -d\tau^2 + \cosh^2\tau \left(d\theta^2 + \cos^2\theta\, d\Omega^2_{d-2} \right), \label{eq:global_dS_metric} \\
        \text{Poincar\'{e} coordinates:}\quad ds^2 &= \frac{-dz^2+dx_1^2+\cdots+dx_{d-1}^2}{z^2}. \label{eq:Poincare_dS_metric}
    \end{align}
\end{subequations}
Consider the coordinate transformation between the global and Poincar\'e patches of dS$_{d}$. The $d$-dimensional de Sitter spacetime can be embedded in $\mathbb{R}^{1,d}$ as
\begin{equation} \label{eq:dS_hyperboloid}
    - X_0^2 + X_1^2+\cdots+X_d^2 = 1.
\end{equation}
For the global patch introduced in equation \eqref{eq:global_dS_metric}, the embedding is given by
\begin{align}
    \begin{split}
        X_0 &= \sinh \tau, \\
        X_1 &= \cosh \tau\,\sin\theta, \\
        X_2 &= \cosh \tau\,\cos\theta\cos\theta_2, \\
        X_3 &= \cosh \tau\,\cos\theta\sin\theta_2\cos\theta_3, \\
        &\hspace{1cm}\vdots \\
        X_d &= \cosh \tau\,\cos\theta \sin\theta_2\cdots \sin\theta_{d-2}.
      \end{split}
\end{align}
The mapping from the de Sitter hyperboloid \eqref{eq:dS_hyperboloid} to the Poincar\'{e} coordinates \eqref{eq:Poincare_dS_metric} is given by
\begin{align}
    \begin{split}
        X_0 &= -\frac{z}{2} \left(1 - \frac{x^2+1}{z^2} \right), \\
        X_1 &= \frac{z}{2} \left(1 + \frac{1-x^2}{z^2} \right), \\
        X_i &= \frac{x_{i-1}}{z},\quad i=2,\ldots,d,
    \end{split}
\end{align}

The codimension-two extremal hypersurface in Poincar\'{e} coordinates is well known
\begin{equation} \label{eq:RT_Poincare}
    x_{d-1}=0,\quad x_1^2 + \cdots+x_{d-2}^2 = z^2 + \ell^2.
\end{equation}
Using the coordinate transformations above, the extremal surface can be written in the global coordinates \ref{eq:global_dS_metric} as
\begin{equation} \label{eq:RT_global}
    \coth \tau\,\sin\theta = \sin\theta_0,\quad \theta_{d-1} = 0,
\end{equation}
while the angles $(\theta_2,...,\theta_{d-2})$ are free. We  introduce $\theta_0$ such that we have we have $\theta \rightarrow \pm\theta_0$ at the boundary $\tau \rightarrow \tau_\infty\to\infty$. The relation between the subregion radius $\ell$ in Poincar\'{e} patch and the opening angle $\theta_{0}$ of the entangling region follows directly from equations \eqref{eq:RT_Poincare} and \eqref{eq:RT_global},
\begin{equation}
    \ell = \frac{\cos\theta_0}{1+\sin\theta_0}.
\label{eq:ell_theta}
\end{equation}
Similarly, the UV cutoffs in the global and Poincar\'e coordinates are related by
\begin{equation}\label{cutoff relation}
   \frac{1}{\delta}
=
\frac{1+\sin\theta_0}{2}\,
e^{\tau_\infty}
-
\frac{1-\sin\theta_0}{2}\,
e^{-\tau_\infty},
\end{equation}
which implies
\begin{equation}
\frac{2\ell}{\delta}
=
\cos\theta_0\,e^{\tau_\infty}
+\mathcal{O}\!\left(e^{-\tau_\infty}\right).
\label{eq:cutoff_final}
\end{equation}

Let us compute the entanglement entropy from the area of an extremal hypersurface which is stretched between $\theta = -\theta_{0}$ to $\theta = \theta_{0}$ on the sphere $S^{d-1}$ at the asymptotic boundary of the $\left(d+1 \right)$-dimensional wedge geometry. Due to the warped nature of the $\left(d+1 \right)$-dimensional metric \eqref{eq:metric_wedge_for_HEE}, the total area functional assumes the structure
\begin{equation}\label{EE in ds wedge}
    S_{A} = \frac{\text{Area}(\Gamma_{A})}{4G_{N}}=\frac{1}{4G_{N}}\int_{\rho_{1}}^{\rho_{2}} d\rho\, \sinh^{d-2}\rho\, A \left(\gamma_{A}^{\mathrm{dS}} \right),
\end{equation}
where $A \left(\gamma_{A}^{\mathrm{dS}} \right)$ is the area of the extremal hypersurface \eqref{eq:RT_Poincare} or \eqref{eq:RT_global} in a constant $\rho$ de Sitter slice. We do the computation using the Poincar\'{e} coordinates, and later transform to the original, global patch using the relations described above. 

The area of the extremal surface $\gamma_{A}^{\mathrm{dS}}$ is given by integral
\begin{equation}
    A \left(\gamma_{A}^{\mathrm{dS}} \right) = i\, \text{Vol}(S^{d-3}) \int_{\delta/\ell}^{\infty}dy \frac{(1+y^2)^{\frac{d-4}{2}}}{y^{d-2}},\quad y = \frac{z}{\ell},
\end{equation}
where $\delta$ and $\ell$ is related to the cut-off $\tau_{\infty}$ and $\theta_{0}$ via \eqref{cutoff relation}. We compute the area of the extremal surface for half of the Lorentzian dS$_{d}$ i.e. $\tau \geq 0$. To recover the holographic entanglement entropy for the full wedge $-\tau_{\infty} \leq \tau \leq \tau_{\infty}$, we double the result.

\noindent For $d=$ 3, 4, and 5 we obtain the following results
\begin{subequations}
    \begin{align}
        \left. A \left(\gamma_{A}^{\mathrm{dS}} \right) \right|_{d=3} &= 2i\log \frac{2\ell}{\delta} = 2i\log(e^{\tau_{\infty}}\cos\theta_{0}), \label{eq:dS_extsurface_area_3D} \\
        \left. A \left(\gamma_{A}^{\mathrm{dS}} \right) \right|_{d=4} &= 2i\pi \frac{\ell}{\delta}=\pi ie^{\tau_{\infty}}\cos\theta_{0}, \label{eq:dS_extsurface_area_4D} \\
        \left. A \left(\gamma_{A}^{\mathrm{dS}} \right) \right|_{d=5} &= 4\pi i \left(\frac{\ell^2}{2\delta^2}+\frac{1}{2}\log\frac{2\ell}{\delta}+\frac{1}{4} \right) = \frac{\pi i}{2}\cos^2\theta_{0}\, e^{2\tau_{\infty}} + 2\pi i \log \left({\cos\theta_{0}}\,e^{\tau_{\infty}} \right) + \pi i. \label{eq:dS_extsurface_area_5D}
    \end{align}
\end{subequations}

Using equations \eqref{eq:dS_extsurface_area_3D}-\eqref{eq:dS_extsurface_area_5D} in conjunction with \eqref{EE in ds wedge}, we obtain the results in section \ref{sec:HEE}.

\section{Evaluation of the RT Surface Area in the No-island Phase}
\label{appendix:NoIslandArea}

In this appendix, we present the detailed evaluation of the RT surface area in the no-island phase. The extremal surface is anchored on the spherical boundary subregion defined by the angular interval $\theta\in[-\theta_0,\theta_0]$ at the asymptotic AdS boundary.
\subsection*{Area Functional}

On a constant global-time slice, the AdS$_{d+1}$ metric \eqref{eq:ads_global_metric_appendix} reduces to
\begin{equation}
ds^{2}
=
d\rho_g^{2}
+
\sinh^{2}\rho_g
\left(
d\theta^{2}
+
\cos^{2}\theta\,d\Omega_{d-2}^{2}
\right).
\end{equation}

Parameterizing the RT surface by $\rho_g=\rho_g(\theta)$, the induced metric leads to the following area functional:

\begin{equation}
A
=
\Omega_{d-2}
\int d\theta\,
\left(
\sinh \rho_g\,\cos\theta
\right)^{d-2}
\sqrt{
\rho_g'^2+\sinh^{2}\rho_g
},
\label{eq:area_functional_app}
\end{equation}

where

\begin{equation}
\Omega_{d-2}
=
\frac{2\pi^{\frac{d-1}{2}}}
{\Gamma\!\left(\frac{d-1}{2}\right)}
\end{equation}

is the volume of the unit $(d-2)$-sphere.

\subsection*{Evaluation of the Area}

The exact RT surface corresponding to a spherical entangling region, derived in Appendix~\ref{app:RTsurface} (see Equation \eqref{RT_master}), is

\begin{equation}
\tanh \rho_g\,\cos\theta
=
\cos\theta_0.
\label{eq:RT_solution_app}
\end{equation}

From equation\eqref{eq:RT_solution_app}, one finds

\begin{equation}
\sinh \rho_g
=
\frac{\cos\theta_0}
{\sqrt{\cos^{2}\theta-\cos^{2}\theta_0}},
\end{equation}

and

\begin{equation}
\rho_g'
=
\frac{\cos\theta_0\,\sin\theta}
{\cos^{2}\theta-\cos^{2}\theta_0}.
\end{equation}

These relations imply

\begin{equation}
\sqrt{\rho_g'^2+\sinh^{2}\rho_g}
=
\frac{\cos\theta_0\,\sin\theta_0}
{\cos^{2}\theta-\cos^{2}\theta_0}.
\end{equation}
To regulate the UV divergence near the AdS boundary, we introduce an angular cutoff $\epsilon$, so that the integration limits are shifted from $\theta=\pm\theta_0$ to $\theta=\pm(\theta_0-\epsilon)$.Substituting these expressions into Equation \eqref{eq:area_functional_app} and using the reflection symmetry of the RT surface about $\theta=0$, the area reduces to
\begin{equation}
A
=
2\Omega_{d-2}
\cot^{d-1}\theta_0
\int_{0}^{\theta_0-\epsilon}
d\theta\,
\frac{\cos^{d-2}\theta}
{\left(
1-\sin^{2}\theta\,\csc^{2}\theta_0
\right)^{d/2}}.
\label{eq:area_reduced}
\end{equation}

Introducing the change of variables

\begin{equation}
x=\sin^{2}\theta,
\end{equation}

the integral becomes

\begin{equation}
A
=
\Omega_{d-2}
\cot^{d-1}\theta_0
\int_{0}^{\sin^{2}(\theta_0-\epsilon)}
dx\,
x^{-1/2}
(1-x)^{\frac{d-3}{2}}
\left(
1-x\,\csc^{2}\theta_0
\right)^{-d/2}.
\end{equation}

Using the standard integral representation of the Appell hypergeometric function $F_1$ \cite[\href{https://dlmf.nist.gov/16.13}{§16.15}]{NIST:DLMF}, we obtain

\begin{align}
A
=
2\Omega_{d-2}
\cot^{d-1}\theta_0
\sin(\theta_0-\epsilon)
F_{1}
\Bigg(
\frac12,
\frac{3-d}{2},
\frac d2;
\frac32;
\nonumber
\sin^{2}(\theta_0-\epsilon),
\sin^{2}(\theta_0-\epsilon)\csc^{2}\theta_0
\Bigg).
\label{eq:final_area_Noisland}
\end{align}

\section{Holographic stress-energy tensor in dS/CFT} \label{appendix:enmomtensor}

To relate the metric perturbation on the de Sitter brane to the expectation value of the dual stress-energy tensor, we employ the standard dS/CFT dictionary \cite{Strominger:2001pn,Das:2013mfa,McFadden:2009fg}. Near future infinity $\mathcal I^{+}$, an asymptotically de Sitter spacetime can be written in Poincar\'e coordinates as \cite{Starobinsky:1982mr,Skenderis:2002wp}
\begin{equation}
ds^{2}
=
\frac{1}{z^{2}}
\left(
-dz^{2}
+\bar g_{ij}(z,x)\,dx^{i}dx^{j}
\right),
\qquad
z\rightarrow0 .
\end{equation}
The metric admits the asymptotic expansion
\begin{equation}
\bar g_{ij}(z,x)
=
g^{(0)}_{ij}(x)
+z^{2}g^{(2)}_{ij}(x)
+\cdots
+z^{d-1}g^{(d-1)}_{ij}(x)
+\cdots ,
\label{Starobinsky_expansion}
\end{equation}
where $g^{(0)}_{ij}$ is the boundary metric and $g^{(d-1)}_{ij}$ is the normalizable coefficient. The induced metric on a constant-$z$ hypersurface is
\begin{equation}
\gamma_{ij}
=
\frac{1}{z^{2}}
\bar g_{ij}.
\end{equation}

The renormalized Brown--York stress tensor is defined by
\begin{equation}
T^{\rm BY}_{ij}
=
\frac{2}{\sqrt{\gamma}}
\frac{\delta S_{\rm ren}}
{\delta \gamma^{ij}},
\end{equation}
where $S_{\rm ren}$ is the renormalized gravitational action including the Gibbons--Hawking term and the required holographic counterterms \cite{Balasubramanian:1999re,deHaro:2000vlm,Skenderis:2002wp}. Solving the Einstein equations order by order in the asymptotic FG expansion and performing holographic renormalization, one finds
\begin{equation}
T^{\rm BY}_{ij}
=
\frac{d-1}{16\pi G_N^{(d)}}
\,g^{(d-1)}_{ij}.
\label{BYtensor}
\end{equation}

According to the dS/CFT correspondence, the late-time Hartle--Hawking
wavefunction of an asymptotically de Sitter spacetime is identified with
the generating functional of the dual Euclidean CFT\cite{Maldacena:2002vr},
\begin{equation}
Z_{\rm CFT}[g^{(0)}]
=
\Psi_{\rm dS}[g^{(0)}]
\sim
e^{\,iS_{\rm ren}},
\end{equation}
which implies
\begin{equation}
\ln Z_{\rm CFT}
=
iS_{\rm ren}.
\end{equation}
Throughout this work we adopt the convention
$Z_{\rm CFT}=\Psi_{\rm dS}\sim e^{\,iS_{\rm ren}}$.
With this convention, the expectation value of the CFT stress tensor
differs from the renormalized Brown-York tensor by an overall factor of
$i$.The expectation value of the dual stress-energy tensor is therefore
\begin{equation}
\langle T_{ij}\rangle
=
\frac{2}{\sqrt{g^{(0)}}}
\frac{\delta\ln Z_{\rm CFT}}
{\delta g^{(0)ij}}
=
i
\frac{2}{\sqrt{g^{(0)}}}
\frac{\delta S_{\rm ren}}
{\delta g^{(0)ij}}
=
i\,T^{\rm BY}_{ij}.
\end{equation}
Substituting equation~\eqref{BYtensor}, we finally obtain
\begin{equation}
\langle T_{ij}\rangle
=
i\,
\frac{d-1}{16\pi G_N^{(d)}}
\,g^{(d-1)}_{ij}.
\label{dSCFTstress}
\end{equation}

\bibliographystyle{JHEP}
\bibliography{refe}

\providecommand{\href}[2]{#2}\begingroup\raggedright\begin{thebibliography}{10}

\bibitem{Maldacena:1997re}
J.M.~Maldacena, \emph{{The Large $N$ limit of superconformal field theories and supergravity}}, \href{https://doi.org/10.4310/ATMP.1998.v2.n2.a1}{\emph{Adv. Theor. Math. Phys.} {\bfseries 2} (1998) 231} [\href{https://arxiv.org/abs/hep-th/9711200}{{\ttfamily hep-th/9711200}}].

\bibitem{Gubser:1998bc}
S.S.~Gubser, I.R.~Klebanov and A.M.~Polyakov, \emph{{Gauge theory correlators from noncritical string theory}}, \href{https://doi.org/10.1016/S0370-2693(98)00377-3}{\emph{Phys. Lett. B} {\bfseries 428} (1998) 105} [\href{https://arxiv.org/abs/hep-th/9802109}{{\ttfamily hep-th/9802109}}].

\bibitem{Witten:1998qj}
E.~Witten, \emph{{Anti de Sitter space and holography}}, \href{https://doi.org/10.4310/ATMP.1998.v2.n2.a2}{\emph{Adv. Theor. Math. Phys.} {\bfseries 2} (1998) 253} [\href{https://arxiv.org/abs/hep-th/9802150}{{\ttfamily hep-th/9802150}}].

\bibitem{Ryu:2006bv}
S.~Ryu and T.~Takayanagi, \emph{{Holographic derivation of entanglement entropy from AdS/CFT}}, \href{https://doi.org/10.1103/PhysRevLett.96.181602}{\emph{Phys. Rev. Lett.} {\bfseries 96} (2006) 181602} [\href{https://arxiv.org/abs/hep-th/0603001}{{\ttfamily hep-th/0603001}}].

\bibitem{Ryu:2006ef}
S.~Ryu and T.~Takayanagi, \emph{{Aspects of Holographic Entanglement Entropy}}, \href{https://doi.org/10.1088/1126-6708/2006/08/045}{\emph{JHEP} {\bfseries 08} (2006) 045} [\href{https://arxiv.org/abs/hep-th/0605073}{{\ttfamily hep-th/0605073}}].

\bibitem{Hubeny:2007xt}
V.E.~Hubeny, M.~Rangamani and T.~Takayanagi, \emph{{A Covariant holographic entanglement entropy proposal}}, \href{https://doi.org/10.1088/1126-6708/2007/07/062}{\emph{JHEP} {\bfseries 07} (2007) 062} [\href{https://arxiv.org/abs/0705.0016}{{\ttfamily 0705.0016}}].

\bibitem{Hamilton:2006az}
A.~Hamilton, D.N.~Kabat, G.~Lifschytz and D.A.~Lowe, \emph{{Holographic representation of local bulk operators}}, \href{https://doi.org/10.1103/PhysRevD.74.066009}{\emph{Phys. Rev. D} {\bfseries 74} (2006) 066009} [\href{https://arxiv.org/abs/hep-th/0606141}{{\ttfamily hep-th/0606141}}].

\bibitem{Lashkari:2013koa}
N.~Lashkari, M.B.~McDermott and M.~Van~Raamsdonk, \emph{{Gravitational dynamics from entanglement `thermodynamics'}}, \href{https://doi.org/10.1007/JHEP04(2014)195}{\emph{JHEP} {\bfseries 04} (2014) 195} [\href{https://arxiv.org/abs/1308.3716}{{\ttfamily 1308.3716}}].

\bibitem{Faulkner:2013ica}
T.~Faulkner, M.~Guica, T.~Hartman, R.C.~Myers and M.~Van~Raamsdonk, \emph{{Gravitation from Entanglement in Holographic CFTs}}, \href{https://doi.org/10.1007/JHEP03(2014)051}{\emph{JHEP} {\bfseries 03} (2014) 051} [\href{https://arxiv.org/abs/1312.7856}{{\ttfamily 1312.7856}}].

\bibitem{Swingle:2014uza}
B.~Swingle and M.~Van~Raamsdonk, \emph{{Universality of Gravity from Entanglement}},  \href{https://arxiv.org/abs/1405.2933}{{\ttfamily 1405.2933}}.

\bibitem{Dong:2016eik}
X.~Dong, D.~Harlow and A.C.~Wall, \emph{{Reconstruction of Bulk Operators within the Entanglement Wedge in Gauge-Gravity Duality}}, \href{https://doi.org/10.1103/PhysRevLett.117.021601}{\emph{Phys. Rev. Lett.} {\bfseries 117} (2016) 021601} [\href{https://arxiv.org/abs/1601.05416}{{\ttfamily 1601.05416}}].

\bibitem{Faulkner:2017tkh}
T.~Faulkner, F.M.~Haehl, E.~Hijano, O.~Parrikar, C.~Rabideau and M.~Van~Raamsdonk, \emph{{Nonlinear Gravity from Entanglement in Conformal Field Theories}}, \href{https://doi.org/10.1007/JHEP08(2017)057}{\emph{JHEP} {\bfseries 08} (2017) 057} [\href{https://arxiv.org/abs/1705.03026}{{\ttfamily 1705.03026}}].

\bibitem{Akal:2020wfl}
I.~Akal, Y.~Kusuki, T.~Takayanagi and Z.~Wei, \emph{{Codimension two holography for wedges}}, \href{https://doi.org/10.1103/PhysRevD.102.126007}{\emph{Phys. Rev. D} {\bfseries 102} (2020) 126007} [\href{https://arxiv.org/abs/2007.06800}{{\ttfamily 2007.06800}}].

\bibitem{Miao:2020oey}
R.-X.~Miao, \emph{{An Exact Construction of Codimension two Holography}}, \href{https://doi.org/10.1007/JHEP01(2021)150}{\emph{JHEP} {\bfseries 01} (2021) 150} [\href{https://arxiv.org/abs/2009.06263}{{\ttfamily 2009.06263}}].

\bibitem{Miao:2021ual}
R.-X.~Miao, \emph{{Codimension-n holography for cones}}, \href{https://doi.org/10.1103/PhysRevD.104.086031}{\emph{Phys. Rev. D} {\bfseries 104} (2021) 086031} [\href{https://arxiv.org/abs/2101.10031}{{\ttfamily 2101.10031}}].

\bibitem{Geng:2022tfc}
H.~Geng, \emph{{Aspects of AdS$_{2}$ quantum gravity and the Karch-Randall braneworld}}, \href{https://doi.org/10.1007/JHEP09(2022)024}{\emph{JHEP} {\bfseries 09} (2022) 024} [\href{https://arxiv.org/abs/2206.11277}{{\ttfamily 2206.11277}}].

\bibitem{Geng:2022slq}
H.~Geng, A.~Karch, C.~Perez-Pardavila, S.~Raju, L.~Randall, M.~Riojas et~al., \emph{{Jackiw-Teitelboim Gravity from the Karch-Randall Braneworld}}, \href{https://doi.org/10.1103/PhysRevLett.129.231601}{\emph{Phys. Rev. Lett.} {\bfseries 129} (2022) 231601} [\href{https://arxiv.org/abs/2206.04695}{{\ttfamily 2206.04695}}].

\bibitem{Ogawa:2022fhy}
N.~Ogawa, T.~Takayanagi, T.~Tsuda and T.~Waki, \emph{{Wedge holography in flat space and celestial holography}}, \href{https://doi.org/10.1103/PhysRevD.107.026001}{\emph{Phys. Rev. D} {\bfseries 107} (2023) 026001} [\href{https://arxiv.org/abs/2207.06735}{{\ttfamily 2207.06735}}].

\bibitem{Yadav:2023qfg}
G.~Yadav, \emph{{Multiverse in Karch-Randall Braneworld}}, \href{https://doi.org/10.1007/JHEP03(2023)103}{\emph{JHEP} {\bfseries 03} (2023) 103} [\href{https://arxiv.org/abs/2301.06151}{{\ttfamily 2301.06151}}].

\bibitem{Hu:2022lxl}
P.-J.~Hu and R.-X.~Miao, \emph{{Effective action, spectrum and first law of wedge holography}}, \href{https://doi.org/10.1007/JHEP03(2022)145}{\emph{JHEP} {\bfseries 03} (2022) 145} [\href{https://arxiv.org/abs/2201.02014}{{\ttfamily 2201.02014}}].

\bibitem{Miao:2023unv}
R.-X.~Miao, \emph{{Entanglement island and Page curve in wedge holography}}, \href{https://doi.org/10.1007/JHEP03(2023)214}{\emph{JHEP} {\bfseries 03} (2023) 214} [\href{https://arxiv.org/abs/2301.06285}{{\ttfamily 2301.06285}}].

\bibitem{Aguilar-Gutierrez:2023tic}
S.E.~Aguilar-Gutierrez, A.K.~Patra and J.F.~Pedraza, \emph{{Entangled universes in dS wedge holography}}, \href{https://doi.org/10.1007/JHEP10(2023)156}{\emph{JHEP} {\bfseries 10} (2023) 156} [\href{https://arxiv.org/abs/2308.05666}{{\ttfamily 2308.05666}}].

\bibitem{Strominger:2001pn}
A.~Strominger, \emph{{The dS / CFT correspondence}}, \href{https://doi.org/10.1088/1126-6708/2001/10/034}{\emph{JHEP} {\bfseries 10} (2001) 034} [\href{https://arxiv.org/abs/hep-th/0106113}{{\ttfamily hep-th/0106113}}].

\bibitem{Maldacena:2002vr}
J.M.~Maldacena, \emph{{Non-Gaussian features of primordial fluctuations in single field inflationary models}}, \href{https://doi.org/10.1088/1126-6708/2003/05/013}{\emph{JHEP} {\bfseries 05} (2003) 013} [\href{https://arxiv.org/abs/astro-ph/0210603}{{\ttfamily astro-ph/0210603}}].

\bibitem{Blanco:2013joa}
D.D.~Blanco, H.~Casini, L.-Y.~Hung and R.C.~Myers, \emph{{Relative Entropy and Holography}}, \href{https://doi.org/10.1007/JHEP08(2013)060}{\emph{JHEP} {\bfseries 08} (2013) 060} [\href{https://arxiv.org/abs/1305.3182}{{\ttfamily 1305.3182}}].

\bibitem{Wong:2013gua}
G.~Wong, I.~Klich, L.A.~Pando~Zayas and D.~Vaman, \emph{{Entanglement Temperature and Entanglement Entropy of Excited States}}, \href{https://doi.org/10.1007/JHEP12(2013)020}{\emph{JHEP} {\bfseries 12} (2013) 020} [\href{https://arxiv.org/abs/1305.3291}{{\ttfamily 1305.3291}}].

\bibitem{Bhattacharya:2012mi}
J.~Bhattacharya, M.~Nozaki, T.~Takayanagi and T.~Ugajin, \emph{{Thermodynamical Property of Entanglement Entropy for Excited States}}, \href{https://doi.org/10.1103/PhysRevLett.110.091602}{\emph{Phys. Rev. Lett.} {\bfseries 110} (2013) 091602} [\href{https://arxiv.org/abs/1212.1164}{{\ttfamily 1212.1164}}].

\bibitem{Allahbakhshi:2013rda}
D.~Allahbakhshi, M.~Alishahiha and A.~Naseh, \emph{{Entanglement Thermodynamics}}, \href{https://doi.org/10.1007/JHEP08(2013)102}{\emph{JHEP} {\bfseries 08} (2013) 102} [\href{https://arxiv.org/abs/1305.2728}{{\ttfamily 1305.2728}}].

\bibitem{Randall:1999ee}
L.~Randall and R.~Sundrum, \emph{{A Large mass hierarchy from a small extra dimension}}, \href{https://doi.org/10.1103/PhysRevLett.83.3370}{\emph{Phys. Rev. Lett.} {\bfseries 83} (1999) 3370} [\href{https://arxiv.org/abs/hep-ph/9905221}{{\ttfamily hep-ph/9905221}}].

\bibitem{Randall:1999vf}
L.~Randall and R.~Sundrum, \emph{{An Alternative to compactification}}, \href{https://doi.org/10.1103/PhysRevLett.83.4690}{\emph{Phys. Rev. Lett.} {\bfseries 83} (1999) 4690} [\href{https://arxiv.org/abs/hep-th/9906064}{{\ttfamily hep-th/9906064}}].

\bibitem{Heemskerk:2010hk}
I.~Heemskerk and J.~Polchinski, \emph{{Holographic and Wilsonian Renormalization Groups}}, \href{https://doi.org/10.1007/JHEP06(2011)031}{\emph{JHEP} {\bfseries 06} (2011) 031} [\href{https://arxiv.org/abs/1010.1264}{{\ttfamily 1010.1264}}].

\bibitem{Faulkner:2010jy}
T.~Faulkner, H.~Liu and M.~Rangamani, \emph{{Integrating out geometry: Holographic Wilsonian RG and the membrane paradigm}}, \href{https://doi.org/10.1007/JHEP08(2011)051}{\emph{JHEP} {\bfseries 08} (2011) 051} [\href{https://arxiv.org/abs/1010.4036}{{\ttfamily 1010.4036}}].

\bibitem{Guijosa:2022jdo}
A.~Guijosa, Y.D.~Olivas and J.F.~Pedraza, \emph{{Holographic coarse-graining: correlators from the entanglement wedge and other reduced geometries}}, \href{https://doi.org/10.1007/JHEP08(2022)118}{\emph{JHEP} {\bfseries 08} (2022) 118} [\href{https://arxiv.org/abs/2201.01786}{{\ttfamily 2201.01786}}].

\bibitem{deHaro:2000vlm}
S.~de~Haro, S.N.~Solodukhin and K.~Skenderis, \emph{{Holographic reconstruction of space-time and renormalization in the AdS / CFT correspondence}}, \href{https://doi.org/10.1007/s002200100381}{\emph{Commun. Math. Phys.} {\bfseries 217} (2001) 595} [\href{https://arxiv.org/abs/hep-th/0002230}{{\ttfamily hep-th/0002230}}].

\bibitem{Balasubramanian:1999re}
V.~Balasubramanian and P.~Kraus, \emph{{A Stress tensor for Anti-de Sitter gravity}}, \href{https://doi.org/10.1007/s002200050764}{\emph{Commun. Math. Phys.} {\bfseries 208} (1999) 413} [\href{https://arxiv.org/abs/hep-th/9902121}{{\ttfamily hep-th/9902121}}].

\bibitem{Emparan:1999pm}
R.~Emparan, C.V.~Johnson and R.C.~Myers, \emph{{Surface terms as counterterms in the AdS / CFT correspondence}}, \href{https://doi.org/10.1103/PhysRevD.60.104001}{\emph{Phys. Rev. D} {\bfseries 60} (1999) 104001} [\href{https://arxiv.org/abs/hep-th/9903238}{{\ttfamily hep-th/9903238}}].

\bibitem{Israel:1966rt}
W.~Israel, \emph{{Singular hypersurfaces and thin shells in general relativity}}, \href{https://doi.org/10.1007/BF02710419}{\emph{Nuovo Cim. B} {\bfseries 44S10} (1966) 1}.

\bibitem{Duff:1993wm}
M.J.~Duff, \emph{{Twenty years of the Weyl anomaly}}, \href{https://doi.org/10.1088/0264-9381/11/6/004}{\emph{Class. Quant. Grav.} {\bfseries 11} (1994) 1387} [\href{https://arxiv.org/abs/hep-th/9308075}{{\ttfamily hep-th/9308075}}].

\bibitem{Henningson:1998gx}
M.~Henningson and K.~Skenderis, \emph{{The Holographic Weyl anomaly}}, \href{https://doi.org/10.1088/1126-6708/1998/07/023}{\emph{JHEP} {\bfseries 07} (1998) 023} [\href{https://arxiv.org/abs/hep-th/9806087}{{\ttfamily hep-th/9806087}}].

\bibitem{Karch:2000ct}
A.~Karch and L.~Randall, \emph{{Locally localized gravity}}, \href{https://doi.org/10.1088/1126-6708/2001/05/008}{\emph{JHEP} {\bfseries 05} (2001) 008} [\href{https://arxiv.org/abs/hep-th/0011156}{{\ttfamily hep-th/0011156}}].

\bibitem{Holzhey:1994we}
C.~Holzhey, F.~Larsen and F.~Wilczek, \emph{{Geometric and renormalized entropy in conformal field theory}}, \href{https://doi.org/10.1016/0550-3213(94)90402-2}{\emph{Nucl. Phys. B} {\bfseries 424} (1994) 443} [\href{https://arxiv.org/abs/hep-th/9403108}{{\ttfamily hep-th/9403108}}].

\bibitem{Calabrese:2004eu}
P.~Calabrese and J.L.~Cardy, \emph{{Entanglement entropy and quantum field theory}}, \href{https://doi.org/10.1088/1742-5468/2004/06/P06002}{\emph{J. Stat. Mech.} {\bfseries 0406} (2004) P06002} [\href{https://arxiv.org/abs/hep-th/0405152}{{\ttfamily hep-th/0405152}}].

\bibitem{Hung:2011xb}
L.-Y.~Hung, R.C.~Myers and M.~Smolkin, \emph{{On Holographic Entanglement Entropy and Higher Curvature Gravity}}, \href{https://doi.org/10.1007/JHEP04(2011)025}{\emph{JHEP} {\bfseries 04} (2011) 025} [\href{https://arxiv.org/abs/1101.5813}{{\ttfamily 1101.5813}}].

\bibitem{Wehrl:1978zz}
A.~Wehrl, \emph{{General properties of entropy}}, \href{https://doi.org/10.1103/RevModPhys.50.221}{\emph{Rev. Mod. Phys.} {\bfseries 50} (1978) 221}.

\bibitem{Vedral:2002zz}
V.~Vedral, \emph{{The role of relative entropy in quantum information theory}}, \href{https://doi.org/10.1103/RevModPhys.74.197}{\emph{Rev. Mod. Phys.} {\bfseries 74} (2002) 197} [\href{https://arxiv.org/abs/quant-ph/0102094}{{\ttfamily quant-ph/0102094}}].

\bibitem{Casini:2011kv}
H.~Casini, M.~Huerta and R.C.~Myers, \emph{{Towards a derivation of holographic entanglement entropy}}, \href{https://doi.org/10.1007/JHEP05(2011)036}{\emph{JHEP} {\bfseries 05} (2011) 036} [\href{https://arxiv.org/abs/1102.0440}{{\ttfamily 1102.0440}}].

\bibitem{Paul:2018spp}
P.~Paul and P.~Roy, \emph{{Linearized Einstein{\textquoteright}s equation around pure BTZ from entanglement thermodynamics}}, \href{https://doi.org/10.1007/s10714-019-2636-9}{\emph{Gen. Rel. Grav.} {\bfseries 51} (2019) 155} [\href{https://arxiv.org/abs/1803.06484}{{\ttfamily 1803.06484}}].

\bibitem{Geng:2020fxl}
H.~Geng, A.~Karch, C.~Perez-Pardavila, S.~Raju, L.~Randall, M.~Riojas et~al., \emph{{Information Transfer with a Gravitating Bath}}, \href{https://doi.org/10.21468/SciPostPhys.10.5.103}{\emph{SciPost Phys.} {\bfseries 10} (2021) 103} [\href{https://arxiv.org/abs/2012.04671}{{\ttfamily 2012.04671}}].

\bibitem{Hao:2025ocu}
P.-X.~Hao, N.~Ogawa, T.~Takayanagi and T.~Waki, \emph{{Flat space holography via AdS/BCFT}}, \href{https://doi.org/10.1007/JHEP10(2025)159}{\emph{JHEP} {\bfseries 10} (2025) 159} [\href{https://arxiv.org/abs/2509.00652}{{\ttfamily 2509.00652}}].

\bibitem{Nishioka:2021cxe}
T.~Nishioka, T.~Takayanagi and Y.~Taki, \emph{{Topological pseudo entropy}}, \href{https://doi.org/10.1007/JHEP09(2021)015}{\emph{JHEP} {\bfseries 09} (2021) 015} [\href{https://arxiv.org/abs/2107.01797}{{\ttfamily 2107.01797}}].

\bibitem{Mollabashi:2021xsd}
A.~Mollabashi, N.~Shiba, T.~Takayanagi, K.~Tamaoka and Z.~Wei, \emph{{Aspects of pseudoentropy in field theories}}, \href{https://doi.org/10.1103/PhysRevResearch.3.033254}{\emph{Phys. Rev. Res.} {\bfseries 3} (2021) 033254} [\href{https://arxiv.org/abs/2106.03118}{{\ttfamily 2106.03118}}].

\bibitem{Chen:2023gnh}
Z.~Chen, \emph{{Complex-valued Holographic Pseudo Entropy via Real-time AdS/CFT Correspondence}},  \href{https://arxiv.org/abs/2302.14303}{{\ttfamily 2302.14303}}.

\bibitem{Doi:2022iyj}
K.~Doi, J.~Harper, A.~Mollabashi, T.~Takayanagi and Y.~Taki, \emph{{Pseudoentropy in dS/CFT and Timelike Entanglement Entropy}}, \href{https://doi.org/10.1103/PhysRevLett.130.031601}{\emph{Phys. Rev. Lett.} {\bfseries 130} (2023) 031601} [\href{https://arxiv.org/abs/2210.09457}{{\ttfamily 2210.09457}}].

\bibitem{Jiang:2023loq}
X.~Jiang, P.~Wang, H.~Wu and H.~Yang, \emph{{Timelike entanglement entropy in dS$_{3}$/CFT$_{2}$}}, \href{https://doi.org/10.1007/JHEP08(2023)216}{\emph{JHEP} {\bfseries 08} (2023) 216} [\href{https://arxiv.org/abs/2304.10376}{{\ttfamily 2304.10376}}].

\bibitem{Doi:2023zaf}
K.~Doi, J.~Harper, A.~Mollabashi, T.~Takayanagi and Y.~Taki, \emph{{Timelike entanglement entropy}}, \href{https://doi.org/10.1007/JHEP05(2023)052}{\emph{JHEP} {\bfseries 05} (2023) 052} [\href{https://arxiv.org/abs/2302.11695}{{\ttfamily 2302.11695}}].

\bibitem{NIST:DLMF}
``{\it NIST Digital Library of Mathematical Functions}.'' \url{https://dlmf.nist.gov/}, Release 1.2.7 of 2026-06-15.

\bibitem{Almheiri:2019hni}
A.~Almheiri, R.~Mahajan, J.~Maldacena and Y.~Zhao, \emph{{The Page curve of Hawking radiation from semiclassical geometry}}, \href{https://doi.org/10.1007/JHEP03(2020)149}{\emph{JHEP} {\bfseries 03} (2020) 149} [\href{https://arxiv.org/abs/1908.10996}{{\ttfamily 1908.10996}}].

\bibitem{Penington:2019npb}
G.~Penington, \emph{{Entanglement Wedge Reconstruction and the Information Paradox}}, \href{https://doi.org/10.1007/JHEP09(2020)002}{\emph{JHEP} {\bfseries 09} (2020) 002} [\href{https://arxiv.org/abs/1905.08255}{{\ttfamily 1905.08255}}].

\bibitem{Almheiri:2019psf}
A.~Almheiri, N.~Engelhardt, D.~Marolf and H.~Maxfield, \emph{{The entropy of bulk quantum fields and the entanglement wedge of an evaporating black hole}}, \href{https://doi.org/10.1007/JHEP12(2019)063}{\emph{JHEP} {\bfseries 12} (2019) 063} [\href{https://arxiv.org/abs/1905.08762}{{\ttfamily 1905.08762}}].

\bibitem{Hartman:2020swn}
T.~Hartman, E.~Shaghoulian and A.~Strominger, \emph{{Islands in Asymptotically Flat 2D Gravity}}, \href{https://doi.org/10.1007/JHEP07(2020)022}{\emph{JHEP} {\bfseries 07} (2020) 022} [\href{https://arxiv.org/abs/2004.13857}{{\ttfamily 2004.13857}}].

\bibitem{Hashimoto:2020cas}
K.~Hashimoto, N.~Iizuka and Y.~Matsuo, \emph{{Islands in Schwarzschild black holes}}, \href{https://doi.org/10.1007/JHEP06(2020)085}{\emph{JHEP} {\bfseries 06} (2020) 085} [\href{https://arxiv.org/abs/2004.05863}{{\ttfamily 2004.05863}}].

\bibitem{Nakata:2020luh}
Y.~Nakata, T.~Takayanagi, Y.~Taki, K.~Tamaoka and Z.~Wei, \emph{{New holographic generalization of entanglement entropy}}, \href{https://doi.org/10.1103/PhysRevD.103.026005}{\emph{Phys. Rev. D} {\bfseries 103} (2021) 026005} [\href{https://arxiv.org/abs/2005.13801}{{\ttfamily 2005.13801}}].

\bibitem{Mollabashi:2020yie}
A.~Mollabashi, N.~Shiba, T.~Takayanagi, K.~Tamaoka and Z.~Wei, \emph{{Pseudo Entropy in Free Quantum Field Theories}}, \href{https://doi.org/10.1103/PhysRevLett.126.081601}{\emph{Phys. Rev. Lett.} {\bfseries 126} (2021) 081601} [\href{https://arxiv.org/abs/2011.09648}{{\ttfamily 2011.09648}}].

\bibitem{Ratcliffe:2006bfa}
J.G.~Ratcliffe, \emph{{Foundations of hyperbolic manifolds}}, Springer, New York (2006).

\bibitem{Das:2013mfa}
D.~Das, S.R.~Das and K.~Narayan, \emph{{dS/CFT at uniform energy density and a de Sitter 'bluewall'}}, \href{https://doi.org/10.1007/JHEP04(2014)116}{\emph{JHEP} {\bfseries 04} (2014) 116} [\href{https://arxiv.org/abs/1312.1625}{{\ttfamily 1312.1625}}].

\bibitem{McFadden:2009fg}
P.~McFadden and K.~Skenderis, \emph{{Holography for Cosmology}}, \href{https://doi.org/10.1103/PhysRevD.81.021301}{\emph{Phys. Rev. D} {\bfseries 81} (2010) 021301} [\href{https://arxiv.org/abs/0907.5542}{{\ttfamily 0907.5542}}].

\bibitem{Starobinsky:1982mr}
A.A.~Starobinsky, \emph{{Isotropization of arbitrary cosmological expansion given an effective cosmological constant}}, {\emph{JETP Lett.} {\bfseries 37} (1983) 66}.

\bibitem{Skenderis:2002wp}
K.~Skenderis, \emph{{Lecture notes on holographic renormalization}}, \href{https://doi.org/10.1088/0264-9381/19/22/306}{\emph{Class. Quant. Grav.} {\bfseries 19} (2002) 5849} [\href{https://arxiv.org/abs/hep-th/0209067}{{\ttfamily hep-th/0209067}}].

\end{thebibliography}\endgroup

\end{document}